\documentclass[fleqn,usenatbib]{mnras}

\usepackage{newtxtext}
\usepackage[varvw]{newtxmath}
\usepackage[T1]{fontenc}

\DeclareRobustCommand{\VAN}[3]{#2}
\let\VANthebibliography\thebibliography
\def\thebibliography{\DeclareRobustCommand{\VAN}[3]{##3}\VANthebibliography}

\usepackage{graphicx}	% Including figure files
\usepackage{amsmath}	% Advanced maths commands
\usepackage{multirow}
\usepackage{verbatim}
\usepackage{threeparttable}

\title[Revisiting GRB 060218]{Revisiting GRB 060218: new insights into low-luminosity gamma-ray bursts from a revised shock breakout model}

\author[Irwin \& Hotokezaka]{
Christopher M. Irwin,$^{1}$\thanks{E-mail: irwincm@g.ecc.u-tokyo.ac.jp (CMI)}
Kenta Hotokezaka$^{1}$
\\
$^{1}$Research Center for the Early Universe, Graduate School of Science, The University of Tokyo, Bunkyo, Tokyo 113-0033, Japan
}

\date{Accepted XXX. Received YYY; in original form ZZZ}

\pubyear{\the\year{}}

\begin{document}
\label{firstpage}
\pagerange{\pageref{firstpage}--\pageref{lastpage}}
\maketitle

\begin{abstract}
Despite two decades since the discovery of low-luminosity gamma-ray bursts, their origin remains poorly understood. In events such as GRB 060218, shock breakout from a progenitor with an extended ($10^{13}$--$10^{14}\,$cm), low-mass ($0.01$--$0.1\,\mathrm{M}_\odot$) envelope provides one possible interpretation for the smooth prompt X-ray emission lasting $\sim1000\,$s and the early optical peak at $\sim0.5\,$d. However, current shock breakout models have difficulties explaining the unexpectedly strong optical emission at $\sim100\,$s, the simultaneous presence of blackbody and power-law components in the X-ray spectrum, and the rapid evolution of the peak energy. We suggest that these peculiar features can be explained by a recently realized shock breakout scenario, in which the gas and the radiation are initially out of thermal equilibrium, but they achieve equilibrium on a time-scale faster than the light-crossing time of the envelope. In this non-standard case, due to the effects of light travel time, the observed X-ray spectrum is a multi-temperature blend of blackbody and free-free components.  The free-free emission is spectrally broad, peaking in hard X-rays while also enhancing the early optical signal.  As the system thermalizes, the free-free component quickly evolves toward lower energies, reproducing the observed rapid peak energy decay. To match observations, we find that more than $10^{50}\,$erg must be deposited in the envelope, which may be accomplished by a choked jet. These results strengthen the case for a shock breakout origin of $ll$GRBs, and provide further evidence connecting $ll$GRBs to peculiar progenitors with extended low-mass envelopes.
\end{abstract}


%%%%%%%%%%%%%%%%%%%%%%%%%%%%%%%%%%%%%%%%%%%%%%%%%%

%%%%%%%%%%%%%%%%% BODY OF PAPER %%%%%%%%%%%%%%%%%%

\section{Introduction}
\label{sec:introduction}

Low-luminosity gamma-ray bursts ($ll$GRBs) are a type of peculiar gamma-ray burst (GRB) associated with broad-lined Type Ic supernovae (SNe), which are characterized by their low luminosity, long duration, and soft spectrum \citep{liang,virgili}.  Well-studied examples include GRB 980425/SN 1998bw \citep{galama,iwamoto,kulkarni,pian2,patat,kouveliotou}, GRB 031203/SN 2003lw \citep{malesani,sazonov,soderberg2,thomsen,watson,mazzali2}, GRB 060218/SN 2006aj \citep{campana,liang2,mazzali,pian,soderberg,sollerman,amati,mazzali4,toma}, GRB 100316D/SN 2010bh \citep{cano,chornock,fan2,starling,margutti1}, GRB 171205A/SN 2017iuk \citep{delia,wang18,izzo,urata,maity}, and GRB 201015A \citep{patel,belkin,komesh}.  The recently discovered X-ray transient EP240414a \citep{sun,bright,srivastav,vandalen} also bears similarities to these events, but its origin is still debated.  Although their faint emission makes them difficult to observe, $ll$GRBs are especially interesting from a stellar evolution perspective, because they may suggest a distinct progenitor channel with a volumetric rate roughly 10 times larger than typical long GRBs \citep[e.g.,][]{soderberg,guetta}.  However, the nature of $ll$GRB progenitors and the reason for their relative abundance are not yet well understood.
 
Among the $ll$GRB sample, GRB 060218 and GRB 100316D exhibit additional unusual features which  distinguish them from the rest, possibly implying a significantly different progenitor.  These features include: a smooth light curve lacking the usual temporal variability; a very long duration of $\ga 1000 \,\rm{s}$; an excess of soft X-rays around $\sim 0.1 \,\rm{keV}$ that has been interpreted as a blackbody emission component \citep[e.g.,][]{campana,kaneko,toma}; a sharp drop in the X-ray luminosity by orders of magnitude at the end of the prompt phase; bright optical emission peaking around half a day after the burst; a total radiated energy of $\sim 10^{49} \,\rm{erg}$, which is much smaller than the associated SN energy; a radio afterglow indicating mildly relativistic ejecta \citep[e.g,][]{soderberg}; and a soft X-ray afterglow which is unusually bright compared to the radio afterglow \citep[e.g.,][]{barniolduran}, implies an atypically large intrinsic absorption column \citep{margutti2}, and shows possible evidence for chromatic evolution \citep[e.g.,][]{amati,ic16}.  

A comparison of the peculiar properties of GRB 060218 to several other well-observed $ll$GRBs and the X-ray transient EP 240414a is provided in Table~\ref{table1}.  The data are compiled from the references given in the first paragraph, along with \citet{kaneko}, \citet{barniolduran}, and \citet{margutti2}.  GRB 201015A is omitted due to a less complete dataset.  Items labelled with a question mark in the table are inconclusive based on available data.  We see that apart from the unusually low SN energy in GRB 060218 \citep[$E_{\rm SN} \approx 2 \times 10^{51}\,$erg,][]{mazzali}, GRB 060218 and GRB 100316D share many of the same unusual features.  The rest of the sample are more diverse.  While most of the features highlighted in Table~\ref{table1} are present in some objects but not in others, we identify three features which may be universal among these events: a peak energy evolution steeper than $t^{-1}$, a steep drop in the X-ray flux after the prompt phase, and an early peak in the optical flux on a time scale of hours to days.

\begin{table*}
\centering
\begin{threeparttable}
\begin{tabular}{l|c c c c c c}
Event & GRB 060218 & GRB 980425 & GRB 031203 & GRB 100316D & GRB 171205A & EP240414a \\
\hline
\multicolumn{7}{c}{Prompt phase} \\
\hline
Smooth single-peaked light curve? & yes & yes & no\tnote{a} & yes & yes & yes \\
Peak energy $E_{\rm p}< 50\,$keV? & yes & no & no & yes & no & yes \\
Duration $T_{90} > 1000\,$s? & yes & no & no & yes & no & no \\ 
Supernova energy $E_{\rm SN} \sim 10^{52}\,$erg? & no & yes & yes & yes & yes & yes \\
Excess soft emission at $\sim 0.1\,$keV? & yes & no & no & yes & yes & yes\tnote{a} \\
$E_{\rm p}$ decays faster than $t^{-1}$? & yes & yes & yes & yes & ?\tnote{c} & ?\tnote{d} \\
X-ray flux drops rapidly when the prompt phase ends? & yes & ?\tnote{d} & ?\tnote{e} & yes & yes & yes \\
\hline
\multicolumn{7}{c}{Afterglow phase} \\
\hline
Optical flux peaks at $t \sim 0.5\,$d? & yes & ?\tnote{f} & ?\tnote{g} & ?\tnote{f} & yes & yes \\
Brighter X-rays than expected from modelling the radio? & yes & yes & no & yes & no & ?\tnote{h} \\
Soft X-ray emission (i.e. $F_\nu \propto \nu^{-\beta_{\rm X}}$ with $\beta_{\rm X} > 2$)? & yes & no & no & yes & no & ?\tnote{h} \\
Intrinsic absorption $N_{\rm H} > 3 \times 10^{21}$? & yes & no & ?\tnote{i} & yes & no & yes\tnote{j} \\
\end{tabular}
\begin{tablenotes}
\item[a] Two smooth pulses were detected in this case \citep[e.g.,][]{kaneko}.
\item[b] In this event, \textit{only} a soft component was detected \citep{sun}.
\item[c] Unclear because the X-ray analysis \citep[e.g.,][]{delia} was not time-resolved.
\item[d] Unknown because the peak may have been below the \textit{Einstein Probe} bands \citep[e.g.,][]{sun}.
\item[e] Consistent with a steep drop \citep[e.g.,][]{watson}, but uncertain due to a gap in coverage between the prompt emission and afterglow.
\item[f] A peak was not observed, but in the earliest detection there was excess emission in at least one band \citep[e.g.,][]{galama,cano}.
\item[g] Unclear because optical photometry was not obtained until $\sim 10\,$d after the burst \citep{malesani}.
\item[h] Detailed X-ray afterglow analysis has not been reported yet.
\item[i] Unknown due a high Galactic absorption along the line of sight \citep[e.g.,][]{watson}.
\item[j] Estimated from the prompt emission rather than the afterglow \citep{sun}.
\end{tablenotes}
\end{threeparttable}
\caption{Peculiar properties of $ll$GRBs.}
\label{table1}
\end{table*}

Theoretical studies of $ll$GRBs have focused significantly on GRB 060218, which remains perhaps the best-observed $ll$GRB due to the extensive multi-wavelength coverage at early times and the detection of an associated supernova, SN 2006aj \citep{campana,mazzali,pian}.  The main reason for the excellent data was GRB 060218's proximity: at a redshift of $z=0.03$ \citep{campana}, the event was sufficiently bright in UV/optical emission to be observed by \textit{Swift}'s UVOT, which was not the case for the more distant event GRB 100316D, located at $z=0.06$ \citep{starling}.  The early UV/optical coverage provides crucial additional constraints on the progenitor properties.  However, because of GRB 060218's many peculiarities, devising a comprehensive model that simultaneously explains all of the observations has proven challenging, despite the rich dataset. 

Thus, even in the best-observed case, our theoretical understanding of $ll$GRBs is lacking.  Broadly speaking, the models proposed so far can be divided into two categories: shock breakout models \citep[e.g.,][]{campana,wang,waxman,li,chevalierfransson,ns2,nakar}, and models involving a central engine \citep[e.g.,][]{toma,ghisellini1,mandal,fan2,thone,ic16,zhang}.  As yet, however, neither picture has provided a fully satisfactory explanation for the multi-wavelength emission of GRB 060218, with the complex multi-component spectrum and the strong optical emission at early times ($\la 1000\,\rm{s}$) being particularly difficult to explain.  An extensive discussion of available models and their drawbacks can be found in \citet{ic16}, and we also provide a brief history of relevant shock breakout models for GRB 060218 in Section~\ref{sec:history} below.  

In this paper, we present a modified shock breakout model which can account for the discrepancies in existing models, providing an improved theoretical picture for the prompt emission of $ll$GRBs.  We begin with a review of the important features of GRB 060218 in Section~\ref{sec:overview}, first discussing the observations in Section~\ref{sec:observations}, and then going over previous shock breakout models for the event and their limitations in Section~\ref{sec:history}.  In  Section~\ref{sec:theory}, we consider the robust conclusions that can be drawn from the data and the literature, and provide some motivation for considering a revised shock breakout model. Next, in Section~\ref{sec:physicalpicture}, we describe the physical picture we have in mind for GRB 060218, first covering the prompt emission in Section~\ref{sec:prompt} and the optical peak in Section~\ref{sec:opticalpeak}, and then discussing the broader $ll$GRB population in \ref{sec:population}.  Finally, in Section~\ref{sec:grb060218}, we present our model for the spectra and light curves (Section~\ref{sec:results}), and consider implications for the  progenitor and explosion properties (Section~\ref{sec:interpretation}). We conclude in Section~\ref{sec:conclusions} with a discussion of what our results imply for future $ll$GRB observations.  

This paper is the third in a three-paper series about shock breakout spectra.  The first paper (Irwin \& Hotokezaka 2024a, hereafter Paper I) describes the technical details of the spectral model used in this work, while the second paper (Irwin \& Hotokezaka 2024b, hereafter Paper II) gives an overview of the possible spectra produced by shock breakout.

\section{overview}
\label{sec:overview}

\subsection{Observational features}
\label{sec:observations}

GRB 060218 was observed by the \textit{Swift} satellite, and detected with the BAT, XRT, and UVOT instruments.  In this paper, we will focus on the prompt X-ray emission lasting until the start of the afterglow phase at $\sim 10^4 \,\rm{s}$, and the early optical emission up to the peak at around $\sim 40000 \,\rm{s}$.  We do not treat the X-ray or radio afterglow, although it is worth noting that the X-ray afterglow in this event was also peculiar \citep[see discussion in, e.g.,][]{barniolduran,margutti2,ic16}.

The joint XRT-BAT spectrum was well-fit at high energies by a cutoff power-law with a peak energy $E_{\rm p} \sim 30 \,\rm{keV}$ \citep{campana,kaneko} and a photon index $\approx -1.4$, but displayed excess emission at energies below $1 \,\rm{keV}$.  Both \citet{campana} and \citet{kaneko} found that including a blackbody component with a temperature of $\sim 0.1\,\rm{keV}$ significantly improved the fit.  \citet{toma} later obtained a better fit to the non-blackbody\footnote{The high-energy power-law component is usually referred to as the `non-thermal component' to distinguish it from the `thermal component' at 0.1\,keV, which can be fitted by a blackbody.  However, we point out this power-law component is not necessarily produced by non-thermal electrons -- it could be due to thermal bremsstrahlung, for instance.  To make this distinction clear, we refer to the two components as the `blackbody' and `non-blackbody' components instead.} component using a Band function model with a comparable peak energy, a photon index $\alpha_{\rm B} \approx -1.1$ below the break, and a photon index $\beta_{\rm B} \approx 2.5$ above the break.  The observed peak energy of the non-blackbody component decreased rapidly with time as $t^{-1.4}$ or $t^{-1.6}$, depending on the spectral fit \citep{kaneko,toma}.  In contrast, the temperature of the blackbody component remained almost constant during the early evolution.

The X-ray light curve was remarkable among GRBs due to its smoothness, long duration, and slow rise.  The integrated $0.3$--$10\,\rm{keV}$ XRT light curve initially rose as $t^{0.6}$ to a peak luminosity of $\sim 3\times 10^{46} \,\rm{erg\,s}^{-1}$ at $\sim 1000\,\rm{s}$, and then fell steeply, eventually dropping below the afterglow flux at around $10^4 \,\rm{s}$ \citep{campana}.  The luminosity of the blackbody component was initially a small fraction of the total XRT luminosity; however, the ratio of the blackbody luminosity to the total XRT luminosity grew over time \citep{campana}.  At around $3000 \,\rm{s}$, the blackbody and non-blackbody luminosities became comparable, and from then on the blackbody component dominated the emission during the steep decay \citep{campana}.  The integrated $15$--$150\,\rm{keV}$ BAT light curve  initially behaved similarly to the XRT light curve, with a comparable luminosity and power-law rise, but it began to decline sooner, at around $500 \,\rm{s}$ \citep{campana,toma}.  

Strong UV/optical emission was detected from early times, rising slowly as roughly $t^{0.4}$ to a peak at $\approx 40000\,\rm{s}$ \citep{campana}.  However, analysis of the UV/optical spectrum is complicated by uncertainties surrounding the host galaxy extinction \citep[see also][for further discussion]{ic16,emery}.  The Galactic reddening along the line of sight is well-constrained, ${E(B-V) = 0.13\text{--}0.14}$ \citep{guenther,sollerman}, but different authors have used different values for the host reddening.    \citet{campana}, \citet{waxman}, \citet{nakar}, and \citet{emery} adopted a relatively high value of $E(B-V) \approx 0.2$, in which case the optical spectrum is consistent with the Rayleigh-Jeans tail of a high temperature blackbody.  A similarly high value of $E(B-V) \ge 0.15$ was estimated by \citet{gorosabel} from polarimetric measurements. On the other hand, Na I D line observations \citep{guenther} suggest a much more modest reddening, $E(B-V) = 0.042$.  Fits to the host galaxy's spectral energy distribution also suggest a low extinction \citep{sollerman}.  Interestingly, a light echo model for the X-ray afterglow considered by \citet{ic16} implied a low extinction as well, consistent with these observations.

Therefore, following \citet{thone} and \citet{ic16}, we adopt the lower extinction value.  By re-analysing the data of \citet{campana} with this lower extinction, \citet{thone} estimated a UV/optical luminosity of $\approx 5\times 10^{43} \,\rm{erg\,s}^{-1}$ and an effective temperature of $\approx 35000\,\rm{K}$ near peak light.  The inferred spectrum in this case is flatter than a Rayleigh-Jeans spectrum, with roughly $F_\nu \propto \nu$ in the optical band at 30\,ks, instead of $F_\nu \propto \nu^2$.

Another peculiar feature of the optical spectrum, which has not received much attention so far, is that the spectral slope seems to increase with time.  This is true regardless of the assumed extinction, and interestingly, it is contrary to the expectation for a constant-temperature (or cooling) blackbody which peaks above the optical band, in which case the spectral index should be constant (or strictly decreasing).  The difference in spectral slope can be seen clearly, e.g., by comparing the optical spectrum at $2000 \,\rm{s}$ to the spectra at later times in Figure 1 of \citet{ghisellini1}.  Investigating the early time data of \citet{campana}, but applying the lower extinction value of \citet{thone}, we find that $F_\nu$ is almost constant with $\nu$ at the earliest times.  An independent analysis by \citet{emery} also found evidence for a steepening spectrum (see their Figure 5), with the slope evolving from $F_\nu \propto \nu^{-0.2}$ at 650\,s to $F_\nu \propto \nu^{1.8}$ at 1625\,s.  All of these results suggest that the spectral index in the optical changed by 1--2 within the first few thousand seconds of the burst.  

The X-ray absorption column obtained from fitting the prompt and afterglow spectra \citep{campana,kaneko,margutti2} is $N_{\rm H}\approx 6\times 10^{21} \,\rm{cm}^{-2}$.  As pointed out by \citet{ic16}, this is not consistent with with the optical/UV extinction inferred by \citet{guenther}, possibly implying that the X-rays are absorbed in the immediate circumstellar environment, where the dust has been evaporated.

\subsection{Shock breakout models: progress and problems}
\label{sec:history}

Motivated by the smooth signal, the relatively soft spectrum, and the low radiative efficiency, many authors have focused on a shock breakout interpretation of the prompt emission in GRB 060218.  Here, we briefly review the history of shock breakout models for this event, the progress made so far, and the remaining problems. 

It is important to note that, although we mainly focus on the shock breakout interpretation in this work, emission mechanisms other than shock breakout have also been explored for GRB 060218 by several authors \citep[e.g.,][]{bjornsson,dai,ghisellini1,wang,mandal,toma,ic16,zhang}.  We stress that our aim here is only to investigate the plausibility of the shock breakout view, and not to rule out the central-engine driven scenario, which remains a plausible alternative. For further discussion of alternative models and how they compare to the shock breakout picture, see Sections 1 and 3 of \citet{ic16}.

Early shock breakout models by \citet{campana} and \citet{waxman} suggested that the early optical emission and the blackbody X-ray component were both produced by a mildly relativistic shock breakout in a stellar wind, with the X-rays produced by the shocked wind, and the optical produced by cooling of the outermost layers of the star.  However, there are several issues with this interpretation \citep[see also][for further discussion]{ghisellini1,ghisellini2,ic16}.  First, the duration of the prompt X-ray emission, $\sim 3000 \,\rm{s}$, is much longer than the light-crossing time in the wind breakout model of \citet{waxman}, which is $\sim 170 \,\rm{s}$ for their breakout radius of $\sim 5 \times 10^{12}\,\rm{cm}$.   Although \citet{waxman} speculated that asymmetry may account for this discrepancy, recent work on non-spherical shock breakouts \citep[e.g.,][]{il} suggests that even under the most optimistic conditions, asymmetry is not enough to account for the difference, and therefore a larger radius is needed.  A second problem is that the early models assumed that the matter and radiation were in thermal equilibrium in the postshock region; however, later works by \citet{katz} and \citet{ns} showed that this is not expected to be the case for mildly relativistic shocks.  This problem also extends to the models of \citet{dai} and \citet{wang}, who assumed that shock breakout provides a source of thermal photons which are then upscattered by a fast outflow to produce the prompt X-ray emission.

Meanwhile, \citet{li} and \citet{chevalierfransson} considered a different scenario, in which the optical emission was powered by a SN shock breakout from a stellar mass progenitor, but the X-ray emission was not.  Their results implied a significantly larger breakout radius of $\sim 5 \times 10^{13} \,\rm{cm}$, indicating that the progenitor could not be a bare WR star.  But, in this case it is not clear what the power source for the X-ray emission could be.

\citet{ns2} later suggested that $ll$GRBs are produced by mildly relativistic shock breakouts powered by SNe, and showed that the observed properties of GRB 060218 roughly obey the predicted closure relations for relativistic breakouts.  Relativistic breakouts also produce a characteristic broken power-law spectrum which could plausibly reproduce the Band-function like spectrum of the non-blackbody X-rays in GRB 060218.  However, one problem with this interpretation is the rather modest energy of SN 2006aj, $2 \times 10^{51} \,\rm{erg}$ \citep{mazzali}.  The relativistic shock breakout model requires $\ga 10^{49} \,\rm{erg}$ to be coupled to relativistic ejecta, which is difficult to achieve with such an ordinary SN energy.  For example, even for a SN energy $10^{52} \,\rm{erg}$, \citet{tan} found that only $\sim 10^{48} \,\rm{erg}$ goes into relativistic material.  Also, as \citet{fs2} recently pointed out based on their updated relativistic breakout model, the breakout temperature in GRB 060218 is likely too low for significant pair production, in which case the closure relations do not apply.  They conclude that GRB 060218 may have been a shock breakout, but if so, it was sub-relativistic.

A more recent breakthrough in our understanding came with the realization that early optical peaks like the one observed in GRB 060218 are actually present in many SNe \citep[see][for a recent review]{modjazarcavi}.  While this feature is most commonly associated with Type IIb SNe, several Type Ibc SNe with early peaks have been discovered as well, including the Type Ib events SN 2008D \citep{modjaz,mazzali3} and SN 2019tsf \citep{zenati}, a Type Ibn SN iPTF13beo \citep{gorbikov}, and a few Type Ic-bl SNe, such as SN 2017iuk/GRB 171205A \citep{izzo}, SN 2012bz/GRB 120422A \citep{schulze}, and SN 2020bvc \citep{izzo2,ho}.  Thanks to the UVOT observations of GRB 060218, SN 2006aj is one of the first and best examples of this type of double-peaked Type Ic-bl SNe.  Curiously, the Type Ic-bl SNe with this feature are often associated with $ll$GRBs (or with an intermediate-luminosity GRB, in the case of GRB 120422A).  Even in SN 2020bvc, where no prompt high-energy emission was detected, the afterglow is strikingly similar to that of GRB 060218 \citep{ho}.  This surprising connection motivates the consideration of a scenario where the peculiar prompt emission of $ll$GRBs and the double-peaked nature of their associated SNe are somehow linked.

The usual interpretation for the early optical peak is that it arises from a non-standard progenitor structure where most of the stellar mass resides in a central core, but a small fraction of the mass is present in an envelope or circumstellar material (CSM) which extends out to a radius much larger than the core's radius \citep[e.g.,][]{np}.  In this scenario, the initially optically thick low-mass envelope is heated by the passage of the supernova shockwave, and radiates its internal energy at UV/optical wavelengths once it expands enough for the trapped radiation to diffuse out.  Applying this analysis to the observed optical peak in GRB 060218 suggests that its progenitor may have had an extended envelope with a mass $\sim 0.01 \, \mathrm{M}_\odot$ and a radius of $\sim 10^{13}$--$10^{14} \rm{\,cm}$ \citep{nakar,ic16}.

The inferred presence of this extended envelope led \citet{nakar} to propose a revised shock breakout model which solves many of the problems with previous models.  In his picture, a typical GRB jet is launched into a progenitor system with a WR-star-like core and a surrounding low-mass envelope.  The jet successfully penetrates the core, but the central engine shuts off before the jet can traverse the envelope, so that the jet is choked in the envelope.  The resulting quasi-spherical shock wave generates shock breakout emission in X-rays when it emerges from the edge of the envelope, and later on, cooling emission from the expanding envelope powers an optical transient.  This scenario resolves the issue of the long duration of the prompt emission, as the duration in this case is set by the light-crossing time of the extended envelope, which could exceed $1000 \,\rm{s}$.  It also alleviates the large energy requirements for producing substantial relativistic ejecta, due to the much smaller mass of the envelope compared to the total stellar mass.  Furthermore, the envelope properties needed to produce a shock breakout with the observed luminosity and duration are in rough agreement with the properties inferred from the early optical peak \citep{nakar}.

Although the model of \citet{nakar} is promising, it still struggles to explain several notable features of GRB 060218.  First, in a standard shock breakout scenario, the origin of separate blackbody and non-blackbody components in the X-ray spectrum, each with a distinct temperature, is not clear.  The breakout is expected to produce a blackbody-like spectrum if the matter and radiation behind the shock achieve thermal equilibrium, and a free-free-like spectrum (possibly altered by Comptonization) if they do not.\footnote{Due to the non-negligible difference in the arrival time of light from different parts of the breakout surface, an observer sees contributions from ejecta with a range of different temperatures at a given time. As a result, the observed spectrum can be modified from a simple blackbody or free-free spectrum (for further discussion, see, e.g., Paper I and Paper II).}  If the system is initially out of equilibrium, it evolves towards equilibrium over time.  Although the spectrum may smoothly evolve from a free-free-like to a blackbody-like spectrum, at any given time, the emission is essentially characterized by one temperature.  In contrast, GRB 060218's X-ray spectrum consists of both blackbody emission peaking at $\sim 0.1\,\rm{keV}$ and non-blackbody emission peaking at $\sim 30 \,\rm{keV}$ \citep{campana,kaneko} throughout most of the prompt phase.  A second problem is the rapid evolution of the peak energy.  The standard shock breakout model predicts an observed temperature which evolves as $\propto t^{-\alpha}$, with $\alpha \approx 1/3\text{--}2/3$ for non-relativistic breakouts, and $\alpha \approx 0.5\text{--}1.0$ for relativistic breakouts \citep{ns,ns2}.  Meanwhile, the peak energy of the non-blackbody component in GRB 060218 displays a more rapid decline, with $\alpha \approx 1.4\text{--}1.6$ \citep{kaneko,toma}.  A final issue is the optical emission observed at very early times.  In the standard model, the initial breakout emission peaks in X-rays, and bright optical emission is not expected until later, after the ejecta have cooled significantly.  While breakout models can reproduce the optical emission around the peak at $\sim 40000\,\rm{s}$, the fact that GRB 060218 was already bright in \textit{V}-band $\sim 100\,\rm{s}$ after detection is difficult to reconcile with standard shock breakout theory.  The early emission in optical is considerably brighter than the Rayleigh-Jeans tail of the $0.1 \, \rm{keV}$ blackbody used to fit the X-ray spectrum, and attempting to join the optical and X-ray spectra with a multi-temperature blackbody implies an unphysically large photospheric radius of $> 10^{15} \,\rm{cm}$ at $2000 \,\rm{s}$ \citet{ghisellini1}, in contradiction with the photospheric velocity of $2\times 10^4 \,\rm{km \, s}^{-1}$ measured by \citet{pian}.  These discrepancies encourage us to reevaluate the standard shock breakout scenario for GRB 060218.

\section{Theoretical considerations: what do we know for sure?}
\label{sec:theory}

Let us now consider a few robust conclusions that can be drawn from the available optical and X-ray data.  The first inevitable conclusion is that, regardless of the details of the breakout, the breakout radius must be much larger than the radius of a typical WR star.  All the breakout models considered in the literature so far share this conclusion.  It is true whether the breakout produced the X-ray emission, the optical emission, or both.  It is necessary for the prompt X-ray emission because the long duration implies a large light-crossing time, and it is necessary for the optical emission in order to avoid large losses due to adiabatic expansion.  Assuming spherical symmetry for simplicity, the light-crossing time constraint implies a radius of $\sim 10^{14} \,\rm{cm}$ for a burst duration of $\sim 3000\,\rm{s}$.

At the same time, because of the fast time-scale of the optical transient, the mass involved in producing the optical emission must be rather low.  Taken together, the low mass and large radius imply an extremely low initial density for the emitting material, ${\rho_{\rm env} \approx 5\times10^{-12} M_{\rm env,-2} R_{\rm env,14}^{-3} \,\rm{g\,cm}^{-3}}$, where ${M_{\rm env,-2} = M_{\rm env}/(10^{-2} \mathrm{M}_\odot)}$, ${R_{\rm env,14} = R_{\rm env}/(10^{14}\,\rm{cm})}$, and following \citet{nakar} and \citet{ic16}, we have assumed the mass $M_{\rm env}$ to be located in an extended envelope, with most of the mass concentrated around the radius $R_{\rm env}$.

A second robust conclusion is that the shock was fast.  Support for this claim comes from the initial peak energy of $\sim 30 \,\rm{keV}$ which, as shown in Paper I, suggests a velocity of $\sim 0.1 c$.  A consistent estimate is obtained from the photospheric velocity of $\approx 0.07\,c$ measured in SN 2006aj \citep{pian}.  The inference of mildly relativistic ejecta from radio observations \citep{soderberg,fan1,barniolduran} is further evidence that some ejecta were accelerated to high velocities.  However, a non-relativistic interpretation of the radio data is also possible \citep{toma}, particularly if the shock sweeps up the gas responsible for the strong X-ray absorption before $2\,$d \citep{ic16}.  Even in the non-relativistic view, the required shock velocity is still an appreciable fraction of the speed of light, $c$.

We note that even if the radio emission requires mildly relativistic velocities, this does not necessarily imply that the shell that generated the breakout emission was also mildly relativistic.  The reason is simply because the radio emission and the shock breakout emission might not be produced by the same ejecta. The isotropic energy of the material producing the radio is somewhat uncertain as it depends on the assumed microphysics \citep[see Section 4.5 of][for further discussion]{ic16}, but for a standard choice of $\epsilon_{\rm e} = 0.1$ and $\epsilon_{\rm B} = 0.1$, where $\epsilon_{\rm e}$ and $\epsilon_{\rm B}$ are respectively the fractions of postshock energy in relativistic electrons and magnetic fields, the inferred kinetic energy is $E_{\rm iso, radio} \sim 10^{48} \,\rm{erg}$ \citep[e.g.,][]{soderberg,toma,ic16}.  On the other hand, the shell powering the prompt emission is expected to have an isotropic kinetic energy at least as large as the isotropic energy radiated during the prompt phase, which is $E_{\rm iso,prompt}\sim 10^{50} \,\rm{erg}$.  \citet{fan1} and \citet{toma} suggested that $\epsilon_{\rm e}$ and $\epsilon_{\rm B}$ must take on smaller values, based on a radiative efficiency argument which demands $E_{\rm iso, radio} \ga E_{\rm iso, prompt}$.  However, this argument relies on the assumption that the same material responsible for producing the prompt emission also generated the radio.  If we relax this assumption, it is plausible that $\epsilon_{\rm e}$ and $\epsilon_{\rm B}$ had typical values and $E_{\rm iso,radio} \ll E_{\rm iso, prompt}$.  In this view, if the prompt emission was produced by shock breakout, then in order to explain the radio, only a tiny fraction of the breakout ejecta would need to be accelerated to velocities several times higher than the typical breakout velocity.   Given the possible presence of a jet, as well as potential asymmetries in the progenitor mass distribution, this does not seem unreasonable.  The detection of linear polarization in SN 2006aj further strengthens the conclusion of an asymmetric explosion \citep{gorosabel}.

Finally, we can robustly conclude that blackbody emission is not sufficient to explain either the X-rays or the optical at early times.  This is obviously the case for the X-rays because of the Band-function-like high-energy spectrum.  However, as noted in Section~\ref{sec:observations}, the spectral slope in the optical at $\la 2000 \,\rm{s}$ seems difficult to explain with blackbody emission as well, perhaps hinting that non-blackbody processes also play a role in shaping the optical/UV spectrum at early times.  It is especially interesting to note that, as discussed in Section~\ref{sec:observations}, when adopting a low value for the extinction, the optical spectrum at $2000 \,\rm{s}$ follows $F_\nu \propto \nu^0$, as would be expected for free-free emission.  Curiously, the low-energy power-law of the Band function describing the non-blackbody X-rays is also consistent with $F_\nu \propto \nu^0$ \citep{toma}.  In fact, as can be inferred from Figure 1 of \citet{ghisellini1}, if the $\nu^0$ behaviour were extended from the soft X-ray peak at $0.1\,$keV down to the optical band, the UV/optical data points would lie nearly along the extrapolation.  A similar conclusion about the non-blackbody nature of the early optical emission was also reached by \citet{emery}, who suggested that the relevant radiation mechanism was synchrotron emission from external shocks.  However, we prefer a shock breakout interpretation where the early optical is produced by free-free emission instead.  In this view, provided that the free-free emission is initially self-absorbed below optical frequencies, an evolution from $F_\nu \propto \nu^0$ to $F_\nu \propto \nu^2$ would naturally be expected as the self-absorption frequency sweeps up through the optical band.

The inferred low density and high velocity in GRB 060218, along with the non-blackbody spectrum, strongly motivate us to consider a breakout scenario where the gas and radiation are not initially in thermal equilibrium, in which case non-blackbody X-rays arise naturally.  Moreover, based on the apparent connection between the optical spectrum and the non-blackbody X-ray spectrum at early times, we are compelled to investigate the case where both of these components have the same origin (i.e., both belong to the same free-free-like spectrum).  In this view, the excess optical emission at early times, which is difficult to explain via cooling envelope emission, has a clear explanation.  The optical emission at later times can still be produced by a cooling envelope, as in previous models.  Our challenges, then, are 1) to account for the faster-than-expected peak energy evolution of the non-blackbody X-ray component; 2) to comprehend why the non-blackbody emission is spread over such a broad frequency range; and 3) to understand the origin of the X-ray blackbody and explain how it can exist alongside the non-blackbody emission.  This will be the focus of the next section.

\section{Physical picture}
\label{sec:physicalpicture}

When several peculiar features appear together, it is natural to consider whether they may be connected.  Here, we propose a scenario which can simultaneously account for the broad non-blackbody emission, the fast peak energy decay, and the soft blackbody X-ray component in GRB 060218.  We first discuss the prompt emission, then briefly address the early optical peak, and finally consider some implications for the broader $ll$GRB population.

\subsection{The prompt emission}
\label{sec:prompt}
As discussed in Section~\ref{sec:observations}, after several thousand seconds the prompt X-ray emission appears to be dominated by the soft blackbody component.  Therefore, the non-blackbody component must fade away during the course of the event.  To reproduce this behaviour, we envision the following scenario.  At sufficiently early times, the emission at all frequencies is dominated by a non-blackbody spectrum which radiates strongly in both optical and X-rays, with a flat spectrum $F_\nu \propto \nu^0$.  This non-blackbody component has a rapidly decaying peak energy, as needed to fit the X-ray observations.  At some point, a blackbody component peaking in soft X-rays develops, which alters the observed X-ray spectrum, but contributes negligibly to the optical emission at early times.  The blackbody component cools slowly, if at all.  Eventually, the non-blackbody emission component fades away, and the blackbody emission becomes dominant.  After the non-blackbody component disappears, the X-ray luminosity remains strong until the blackbody component fades (or until its temperature drops below the XRT band, whichever happens first).  Meanwhile, in the optical, the cooling envelope emission (which, for now, we view as a distinct blackbody component) grows stronger and overtakes the non-blackbody component, causing the optical spectrum to steepen to $F_\nu \propto \nu$.  The optical light curve continues to rise, peaks when the envelope has radiated most of its energy, and then steeply decays.

While an overall progression from non-blackbody to blackbody emission is expected in the standard non-equilibrium shock breakout picture, explaining such a complex multi-component light curve is not straightforward.   It seems clear that a standard shock breakout, which generally produces either a blackbody or free-free spectrum at a given time (but not both), will not suffice.  A way of producing a mixture of blackbody and non-blackbody emission is therefore required.  

In Paper I, we have shown that the smearing of the shock breakout spectrum due to a non-negligible light-crossing time can provide the requisite blending of blackbody and non-blackbody emission.  The necessary conditions to achieve this are: 1) the shock velocity should be sufficiently fast that the gas and radiation at the breakout location are initially out of thermal equilibrium; and 2) the temperature evolution should be fast enough that ejecta in thermal equilibrium are revealed before a time of $\approx R_{\rm env}/c$ has elapsed.  Following Paper I, we refer to breakouts meeting these criteria, which are characterized by the simultaneous presence of blackbody and free-free emission components in their spectrum, as INERT breakouts (`initially non-equilibrium rapid thermalization breakout').  Obtaining an INERT breakout requires a shock velocity comparable to $0.1\,c$ (see Paper I), consistent with the expectation for GRB 060218 (as discussed in Section~\ref{sec:theory}).  In this view, on account of the second requirement, fast temperature evolution and a multi-component spectrum go hand-in-hand.  Furthermore, as we will discuss below, non-equilibrium breakout naturally leads to a broad free-free spectrum, as desired. 

In order to satisfy the criteria for INERT breakout, it is also necessary to have a sufficiently high density at the site of shock breakout (Paper I). As discussed in Section~\ref{sec:theory},  we prefer a scenario like the one envisioned by \citet{nakar}, where the shock breakout takes place at the edge of an extended low-mass envelope.  The envelope density $\rho_{\rm env} \sim 10^{-11}\,$g\,cm$^{-3}$ inferred in Section~\ref{sec:theory} is also roughly compatible with the limit of $\rho_{\rm bo} \ga 4 \times 10^{-12}R_{\rm env,14}^{-15/16}\,\text{g}\,\text{cm}^{-3}$ given for INERT breakout in Paper I, provided that density at the breakout location, $\rho_{\rm bo}$, is not too different from the typical envelope density, $\rho_{\rm env}$. This holds only when the density profile is such that breakout occurs before significant shock acceleration takes place (for further discussion, see Section 3 of Paper I).

Putting it all together, it seems that a non-equilibrium breakout from an extended envelope, smeared by light travel effects, potentially has the necessary ingredients to explain the complex spectrum of GRB 060218.  Here, we will offer a brief, qualitative explanation of how this scenario addresses the three challenges raised at the end of Section~\ref{sec:theory}. Readers interested in further technical details may refer to Paper I.  More quantitative modelling of the spectrum will be carried out later, in Section~\ref{sec:grb060218}. 

\begin{enumerate}

\item \textit{Fast peak energy decay:} During the early planar phase of shock breakout, the mass coordinate $m$ from which the observed radiation originates grows logarithmically with time \citep{fs}.  However, in the case of non-equilibrium shock breakout, the radiation temperature within the expanding flow is a steep function of $m$ \citep[e.g.,][]{fs}.  As a result, even though $m$ grows slowly, the observed temperature can decrease rapidly.  In Paper I, we showed that the observed temperature, $T_{\rm obs}$, can decay as steeply as ${T_{\rm obs} \propto t^{-2.2}}$, with some dependence on the density profile and the degree of Comptonization.  This is more than steep enough to explain the $T_{\rm obs} \propto t^{-1.6}$ behaviour observed in GRB 060218.

\item \textit{Broad free-free emission}: In a non-equilibrium shock breakout, the expected spectrum is a (possibly Comptonized) free-free spectrum, with an observed temperature which depends very strongly on the shock velocity at breakout, $v_{\rm bo}$, as $T_{\rm obs} \propto v_{\rm bo}^8$ \citep[e.g.,][]{ns}.  The self-absorption frequency, $\nu_{\rm a}$ also has a very strong inverse dependence on $v_{\rm bo}$, as $\nu_{\rm a} \propto v_{\rm bo}^{-7}$ (Paper I).  Thus, for fast shocks the spectrum is very naturally spread over a wide frequency range.  In Paper I, we estimate $h\nu_{\rm a} \sim 0.1\,$eV and $kT_{\rm obs} \sim 10\,$keV for typical $ll$GRB parameters, indicating that the free-free spectrum extends from optical to X-rays.

\item \textit{Coexisting blackbody and free-free components:} As discussed in Paper I and Paper II, if the time to achieve thermalization is short compared to the light-crossing time, an observer will simultaneously see blackbody and free-free emission originating from different parts of the breakout surface.  Blackbody emission is observed from regions close to the line of sight, while higher-latitude regions are still seen to produce free-free emission due to their delayed arrival time compared to the line of sight (as illustrated in Figure 1 of Paper I).  After a light-crossing time has elapsed, only blackbody emission is observed.  The overall spectral evolution in this case, which starts out dominated by free-free emission but eventually comes to be dominated by blackbody emission, is similar to what was observed in GRB 060218.
\end{enumerate}

\subsection{Post-breakout evolution and the optical peak}
\label{sec:opticalpeak}

If an explosion drives a shock through an extended optically thick envelope, UV/optical emission is expected as the shocked envelope material expands and cools \citep[e.g.,][]{np}.  The light curve peaks once the system expands enough that the whole envelope mass has been exposed.  Such a model has been applied to successfully explain the properties of the early optical peak in SN 2006aj \citep{nakar,ic16}.  The envelope shock breakout model introduced in Paper I also leads to an optical peak, with the same properties as in the \citet{np} model.  This paints a self-consistent picture where  the unusual properties of the prompt X-ray emission and the peculiar early optical peak are both linked to the presence of an extended low-mass envelope. 

While this idealized picture provides some intuition about how the prompt X-ray emission and the optical peak may be connected, a more careful comparison with observations of GRB 060218 reveals that the situation is more complicated in reality.  Although the time-scale of the optical peak predicted by Paper I is consistent with observations, there are issues with the luminosity and temperature.  There is considerable uncertainty in the bolometric luminosity and the temperature of the optical peak at early times, because the emission peaked above the \textit{Swift} UVOT band, and as discussed in Section~\ref{sec:observations} the extinction to the source is not precisely known.  However, the analysis of \citet[][see their Figure 16]{thone} estimated that the spectrum peaked at $10^{15}\,\text{Hz}$ at 0.5\,d, corresponding to a temperature of $\approx 1\,$eV.  The corresponding bolometric luminosity at that time was a few $10^{43}\,\text{erg}\,\text{s}^{-1}$.  By 1.7\,d, they saw clear evidence for a turnover in the spectrum, which seems to robustly indicate that the temperature dropped below 1 eV by that time.  These values are at odds with Paper I, which predicts a significantly higher temperature and bolometric luminosity at the time of the peak, although the luminosity in the optical band turns out to be similar.

This difficulty of connecting the properties of the prompt emission to the properties of the optical peak was also discussed by \citet{ic16}, who concluded that both components could not be produced by a shock breakout.  The fundamental problem is that shock breakout models predict a slow temperature evolution once the emission has thermalized, roughly as $T_{\rm obs} \propto t^{-1/3}$ in the planar phase, or $t^{-0.6}$ in the spherical phase \citep[e.g.,][]{ns}.  This slow evolution is hard to reconcile with the larger-than-expected difference between the temperature of the prompt blackbody X-rays and the optical peak.  This is illustrated clearly in Figure 17 of \citet{thone}, which seems to show a jump in temperature at around 0.1\,d.  Even if we neglect their data point at 0.1\,d, the observed temperature would have to evolve as $T_{\rm obs} \propto t^{-1.2}$ in order to go from $10^6\,$K at 0.03\,d to $3\times10^4\,$K at 0.5\,d, which is much steeper than can be achieved in the standard shock breakout model.

However, unlike \citet{ic16}, we do not see this issue as a fatal one for the shock breakout interpretation.  The reason is that the standard treatment treats the ejecta as spherically symmetric and homogeneous, and this is an oversimplification.  If the breakout is aspherical or the ejecta inhomogeneous, then the ejecta material could be heated a range of different initial temperatures \citep[for further discussion of aspherical breakout see, e.g.,][]{couch,matzner,suzukishigeyama,afsariardchi,ohtani,linial,il,goldberg}.  Although the material responsible for producing the blackbody X-rays, which is initially hot with a blackbody temperature of $T_{\rm BB,bo} \approx 100\,$eV, might not be able to cool down to $\approx1\,$eV in 1\,d, if an appreciable fraction of the ejecta was only heated to a lower temperature of $T_{\rm BB,bo} \sim 10\,$eV, the temperature of the optical component could perhaps be explained by the cooling of this lower-temperature material.

\subsection{The $ll$GRB population}
\label{sec:population}

As discussed in Section~\ref{sec:prompt} (see also Section 3 of Paper I), special conditions are required in order to produce a shock breakout with properties resembling GRB 060218 (i.e., an INERT breakout); the most constraining of these is that the velocity must lie within a narrow range around $\sim 0.1\,c$.  At the same time, the progenitors of $ll$GRBs may show some diversity, and so it seems unlikely that the necessary conditions could be met in every case.  We therefore stress that in our scenario, $ll$GRBs \textit{will only sometimes exhibit a spectrum with coexisting blackbody and free-free components}.  Indeed, of the $ll$GRBs known so far, only GRB 060218, GRB 100316D, and GRB 171205A show such a spectrum.  Other known $ll$GRBs such as GRB 980425 and GRB 031203 also have a smooth light curve which may be consistent with shock breakout, but they do not show evidence for a blackbody spectral component, instead having only a Band-function or power-law like spectrum.  These events have higher peak energies, implying faster breakout velocities, and seem to be consistent with a mildly relativistic shock breakout \citep{ns2,fs2}.  This would place them deep in the non-equilibrium breakout regime, where a blackbody component is not expected  (for more about the spectrum in this case, see Paper II).  Pair creation may also be relevant for these events \citep[e.g.,][]{ns2,fs2}.

The picture we have in mind is that all $ll$GRBs with a smooth, single-peaked X-ray light curve might be produced by shock breakout from extended envelopes.  In some events, the shock velocity might be faster (indeed even mildly relativistic), so that the breakout spectrum is in the free-free regime and there is no blackbody component. The fast velocity could be due to a lower envelope mass, a larger energy deposited in the envelope, or stronger shock acceleration at the envelope's edge.  In other events, the velocity might be slower, so that the gas and radiation are always in thermal equilibrium, and the spectrum is a blackbody from the beginning.\footnote{So far, there are not any convincing candidates among $ll$GRBs for this type of low-velocity thermal breakout (EP 240414a notably had a very soft spectrum, but its expansion velocity was mildly relativistic, and its connection to $ll$GRBs remains unclear).}  Only in events where the shock velocity is just right (around 10 per cent of $c$) do we expect INERT breakout events in which free-free and blackbody components are present simultaneously, as in GRB 060218.  

In this picture, which class of events is more common depends on the underlying distribution of the event rate with respect to velocity.  On one hand, if the event rate does not depend much on velocity, then INERT breakouts should be less common than mildly relativistic breakouts, because they require the velocity to lie within a certain narrow range.  In this case $ll$GRBs with a blackbody component like GRB 060218 should be relatively uncommon.  On the other hand, if events with faster shocks are intrinsically rarer, then INERT breakouts could be more common than mildly relativistic breakouts, and a significant fraction of $ll$GRBs might exhibit a blackbody feature in their spectrum.  For example, suppose that, based on Paper I, INERT breakout occurs in the proper velocity range $0.05 < \beta\gamma < 0.15$, but mildly relativistic breakout occurs the range $0.15 < \beta\gamma <1$.  Then, if we assume that the event rate per unit $\beta\gamma$ follows a power-law distribution $\mathrm{d}N/\mathrm{d}(\beta\gamma) \propto (\beta\gamma)^{-p}$, the rates of these two types of breakout would be equal for $p \approx 1.4$. 

\begin{figure}
\begin{center}
\includegraphics[width=\columnwidth]{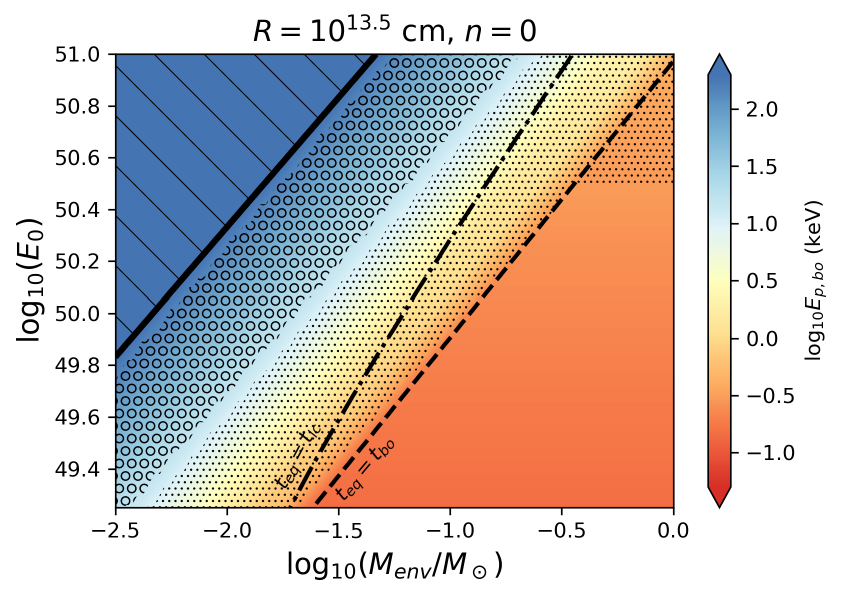}
\end{center}
\caption{The observed peak energy $E_{\rm p,bo} \approx 4 k T_{\rm obs,bo}$ of shock breakout from an extended envelope, for a range of envelope mass and deposited energy.  The hatching with dots and open circles shows where the peak energy lies within the XRT band (0.3--10\,keV) or the BAT band (15--150\,keV), respectively.  An INERT breakout with coexisting free-free and blackbody components in the spectrum is expected in the region between the dashed and dot-dashed lines.  Above the dashed line, the gas and radiation are not in thermal equilibrium initially.  Below the dash-dotted line, equilibrium is achieved in less than a light-crossing time.  For the definitions of $t_{\rm eq}$, $t_{\rm lc}$, and $t_{\rm bo}$, see Section~\ref{sec:grb060218}. Our results are invalid for $kT_{\rm obs,bo} \ga 50\,$keV (indicated by diagonal hatching), because we have neglected pair creation.}
\label{fig:temperaturemap}
\end{figure}

The likelihood of seeing an INERT breakout also depends on the instruments used to detect $ll$GRBs.  Since the observed temperature of non-equilibrium shock breakout depends primarily on shock velocity, the range of velocities for which free-free and blackbody process both contribute can be converted to a range of observed temperatures (or peak energies, assuming $E_{\rm p} \approx 4kT_{\rm obs})$.  To illustrate this, we use the results of Paper I to compute the observed peak energy as a function of the envelope's mass, $M_{\rm env}$, and the energy deposited into it, $E_0$, for a given envelope radius, $R_{\rm env}$, and a power-law index $n$ describing the density profile $\rho \propto (R_{\rm env} -r)^n$.  The result is shown in Fig.~\ref{fig:temperaturemap} for a choice of $n=0$ and $R_{\rm env} = 10^{13.5}\,$cm (see Section 3 of Paper I for further discussion of the $n=0$ case).  We see from this figure that there is significant overlap between the region where INERT breakout is expected (between the dashed and dot-dashed lines), and the region where the peak energy lies in the XRT band (indicated by the dotted hatching).  In this sense, it is not surprising that \textit{Swift}-detected X-ray bright transients like GRB 060218 and GRB 100316D have shown evidence for a blackbody component in their spectra.

\section{A spectral model for GRB 060218}
\label{sec:grb060218}

As suggested in Sections~\ref{sec:physicalpicture}, the INERT breakout scenario explored in Paper I seems to share several features in common with GRB 060218, including a rapid peak energy decay, simultaneous blackbody and power-law emission components, and strong optical emission at early times.  Here, we investigate this connection further by applying the methods of Paper I to observations of GRB 060218.  We first describe the results, then consider implications for the progenitor properties.  Throughout this section, we assume spherical symmetry.

\subsection{Spectra and light curves}
\label{sec:results}

The spectral model of Paper I has six parameters\footnote{GRB 060218 was already bright when it came into \textit{Swift}'s field of view \citep{barbier}, and therefore it is possible that the actual start time of the burst was before the trigger time of \textit{Swift}.  For this reason, we also considered models which included a non-zero delay time between the burst and the \textit{Swift} trigger as an additional parameter.  However, we found that in all cases, a better fit to the data was obtained for small delay times of less than a few hundred seconds. For simplicity, we only present results here for the case where this delay time is fixed to zero.}: the luminosity $L_{\rm bo}$ and observed temperature $T_{\rm obs, bo}$ of the breakout emission, a power-law index $\alpha$ describing the early temperature evolution, the dynamical time of the breakout layer $t_{\rm bo}$, the light-crossing time of the system $t_{\rm lc}$, and the time $t_{\rm eq}$ when blackbody emission is first observed.  We will call these the `spectral parameters.'  These parameters in turn depend on only four parameters describing the progenitor and explosion: the extended envelope's mass $M_{\rm env}$ and radius $R_{\rm env}$, the energy $E_0$ deposited in the envelope, and a power-law index $n$ describing the outermost part of the envelope, which we call the `breakout parameters.'  The spectral parameters are useful, because they have a very direct link to observations (see below). Therefore, we will take the approach of initially treating the six spectral parameters as independent, and then checking later that they are consistent with the four breakout parameters.  Throughout this study, we assume an electron scattering opacity of $\kappa = 0.2\,\text{cm}^2\,\text{g}^{-1}$.

As shown in Paper I, INERT breakout occurs when ${t_{\rm bo} < t_{\rm eq} < t_{\rm lc}}$.  In this case, the breakout duration is set by the light-crossing time, $t_{\rm lc}$.  We will assume for now that this condition is met, and check this later.  Under this assumption, the observed luminosity is related to $L_{\rm bo}$ by $L_{\rm obs} \approx L_{\rm bo} (t_{\rm bo}/t_{\rm lc})$ (see Paper I and Paper II).  The total energy radiated during the breakout is then given by ${L_{\rm obs} t_{\rm lc} \approx L_{\rm bo} t_{\rm bo}}$ (note that in our model, $L_{\rm bo}$ and $t_{\rm bo}$ refer to the luminosity and duration of the emission produced at the source, before smearing over the light-crossing time is taken into account). The model additionally assumes the free-free self-absorption frequency $\nu_{\rm a}$ evolves as $\nu_{\rm a}\propto t^{\omega}$, where ${\omega = \lambda(\alpha -2/3)}$ as derived in Paper I, and $\lambda$ is a function of $n$ with a typical value of $\lambda=1.2\text{--}1.5$. In this study, we adopt $\lambda = 1.5$, which is appropriate for small $n$; we will show in Section~\ref{sec:interpretation} that this is self-consistent.  Lastly, for convenience, we also define the equilibrium temperature, $T_{\rm eq}$, to be the observed temperature at the time $t=t_{\rm eq}$.  

Observations of GRB 060218 immediately place some rough constraints on the spectral parameters.  The burst duration suggests that $t_{\rm lc}\approx 3000\,$s.  The peak XRT luminosity of $\approx 2\times 10^{46}\,\text{erg}\,\text{s}^{-1}$ implies that the total radiated energy of the prompt emission is $L_{\rm bo} t_{\rm bo} \approx 6\times 10^{49}\,$erg. The initial peak energy of $30\,$keV implies an initial temperature of $kT_{\rm obs,bo}\approx 10\,$keV, assuming that the spectrum peaks at $\approx 3kT_{\rm obs}$.  The steeply dropping peak energy at early times ($\propto t^{-1.6}$) suggests $\alpha \approx 1.6$.  The temperatures of the blackbody and non-blackbody components become comparable around the peak of the XRT light curve, suggesting $t_{\rm eq} \approx 2000\,$s.  The temperature around that time is $kT_{\rm eq} \approx 0.1\,$keV. Comparing this to the value of $T_{\rm obs,bo}$ and using $T_{\rm obs} \propto t^{-\alpha}$ with $\alpha\approx 1.6$, we can estimate that $t_{\rm bo} \approx t_{\rm eq} (T_{\rm eq}/T_{\rm obs,bo})^{1/1.6} \approx 100\,$s.  Finally, we combine this with the constraint on the radiated energy to find that $L_{\rm bo}\approx 6\times10^{47}\,\text{erg}\,\text{s}^{-1}$. 

\begin{figure}
\begin{center}
\includegraphics[width=\columnwidth]{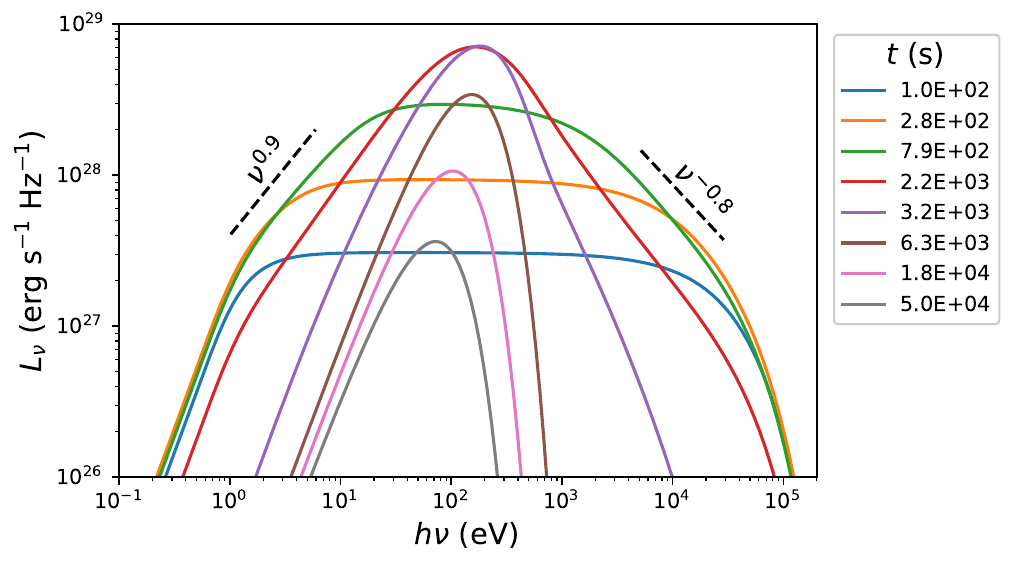}
\end{center}
\caption{Time evolution of the model spectrum for ${L_{\rm bo} = 4 \times 10^{47}\,\text{erg}\,\text{s}^{-1}}$, $T_{\rm obs,bo} = 9$\,keV, $t_{\rm bo} = 100$\,s, $t_{\rm eq} = 2000$\,s, $t_{\rm lc} = 3000$\,s, and $\alpha = 1.6$.  The colour of the lines indicates the time in seconds, as shown in the legend. Dashed lines show the approximate spectral index for optical/UV and X-rays at times $t_{\rm bo} < t < t_{\rm eq}$.}
\label{fig:llgrbspectrumexample}
\end{figure}
   
Motivated by this analysis, we use the method described in Section 4 of Paper I to generate time-dependent model spectra $L_\nu(\nu,t)$, focusing on parameter values close to the rough estimates above.  We integrate these model spectrum over  $0.3$--$10\,$keV and $15$--$150\,$keV to determine the flux in the XRT and BAT bands, respectively.  In addition, we compute the flux in the UVOT instrument's \textit{V}, \textit{B}, \textit{U}, \textit{UVW1}, \textit{UVM1}, and \textit{UVW2} filters. We present here a model with $L_{\rm bo} = 4.0 \times 10^{47}\,\text{erg}\,\text{s}^{-1}$, $t_{\rm lc} = 3000\,$s, $t_{\rm eq} = 2000$\,s, $t_{\rm bo} = 100\,$s, $T_{\rm obs,bo} = 9\,$keV, and $\alpha = 1.6$, which we find gives a satisfactory fit to the data.  Note that for these parameters $t_{\rm bo} < t_{\rm eq} < t_{\rm lc}$, which implies that the spectrum is indeed in the INERT breakout regime.  In order for these six spectral parameters to be compatible with a four-parameter shock breakout interpretation, they must satisfy two additional closure relations: $T_{\rm obs,bo} (t_{\rm bo}/t_{\rm eq})^{[\alpha+\lambda(\alpha-2/3)]/2} \sim (c/a\kappa t_{\rm bo})^{1/4}$ and $L_{\rm bo} t_{\rm bo} \sim 4 \uppi (147 c^{13} k A_{\rm ff}/2 \kappa^5)^{1/4} t_{\rm lc}^2 T_{\rm obs,bo}^{1/8}$, where $A_{\rm ff} = 3.5\times 10^{36}$ in cgs units, as given in Section 4 of Paper I.  Plugging our estimates for the spectral parameters into these equations, we find that the closure relations are satisfied to within a factor of $\sim 2$, indicating that a shock breakout origin is indeed plausible.  The implied values of the breakout parameters will be explored further in Section~\ref{sec:interpretation}.

The overall evolution of the SED in this model is presented in Fig.~\ref{fig:llgrbspectrumexample}.  We see that at early times, the spectrum is a self-absorbed free-free spectrum.  The self-absorption takes place below $\sim 1\,$eV, resulting in considerable early optical emission.  As time goes on, due to the difference in light arrival time across the breakout surface, already-cooling emission from near the line of sight and high-temperature breakout emission from higher latitudes are observed simultaneously.  As a result, at several hundred seconds, a multi-temperature free-free spectrum is observed with a Band function-like broken power-law behaviour.  The energy where the break occurs quickly declines with time.  At $t_{\rm eq} = 2000\,$s, ejecta in thermal equilibrium are exposed near the line of sight, and a blackbody component develops in the spectrum.  Due to its slower temperature evolution, the blackbody component is not significantly smeared out by the effects of light travel time.  At $t\approx t_{\rm lc}=3000\,$s, the free-free component becomes subdominant, and by $t_{\rm lc} + t_{\rm eq} = 5000\,$s, it fades away completely.  After several thousand seconds, the spectrum is a gradually fading and cooling blackbody.

\begin{figure}
\begin{center}
\includegraphics[width =\columnwidth]{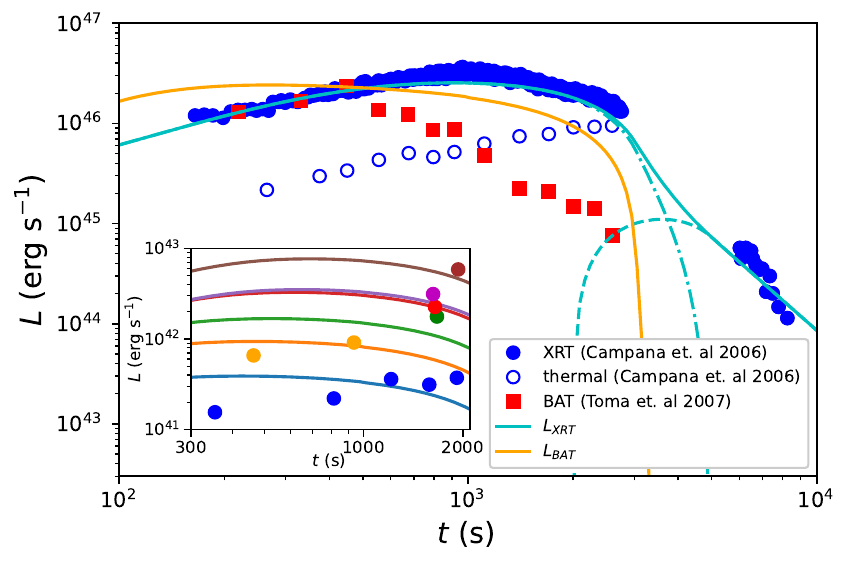}
\end{center}
\caption{Model light curves for XRT (light blue) and BAT (orange), compared to the observations of \citet{campana} (XRT, blue circles) and \citet{toma} (BAT, red squares).  For the XRT light curve, the separate contributions from the blackbody and free-free components are shown with dashed and dot-dashed lines, respectively.  The luminosity of the blackbody component inferred by \citet{campana} is also shown with open circles. \textit{Inset}: Optical light curves during the first $2000\,$s in each UVOT band (\textit{V}, blue; \textit{B}, orange; \textit{U}, green; \textit{UVW1}, red; \textit{UVM1}, purple; \textit{UVW2}, brown), compared to the values reported in \citet{campana} (circles), deabsorbed assuming a reddening (Galactic + host) of $E(B-V) = 0.14 + 0.042$.}
\label{fig:lightcurves}
\end{figure}

Fig.~\ref{fig:lightcurves} compares the light curves (XRT in blue, BAT in orange) generated by the spectrum in Fig.~\ref{fig:llgrbspectrumexample} to observations.  In the figure, the XRT and optical 
measurements are taken from \citet{campana}, while the BAT data are from \citet{toma}.  We focus mainly on fitting the total XRT flux, which as seen in the figure is reproduced very well by the model. In particular, the shape of the XRT light curve is captured accurately, including the slow rise up to the peak and the sharp drop after the peak.  We find that these qualitative features are robust in our model, for a wide range of parameters.  The slow rise to peak occurs as cooler ejecta, with temperatures in the XRT band, are exposed along the long of sight. The sharp drop occurs after $t=t_{\rm lc}$, when even the high-latitude ejecta are observed to be cooling, and their rapidly-falling temperature sweeps through the XRT band.  

\begin{figure}
\begin{center}
\includegraphics[width=\columnwidth]{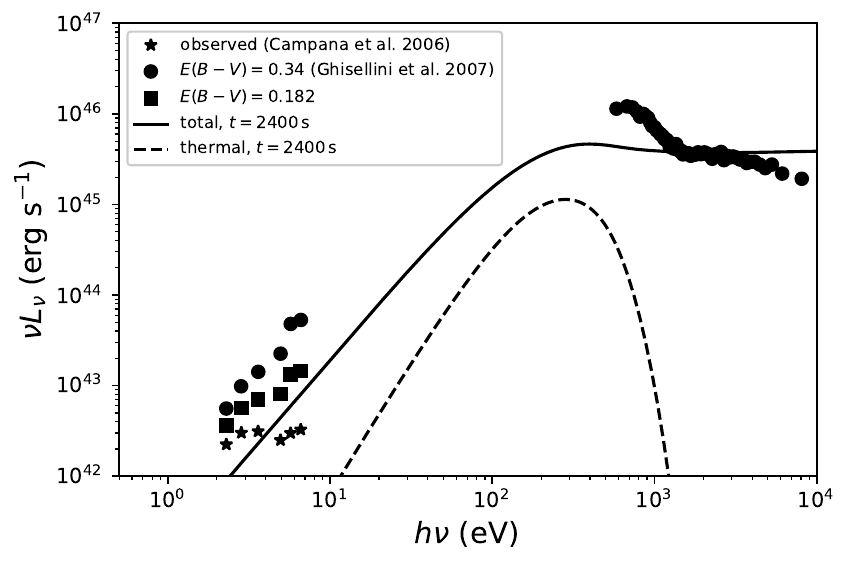}
\end{center}
\caption{The $\nu L_\nu$ spectrum in our model at $2400\,$s (solid), compared to the one presented in Figure 1 of \citet{ghisellini1} (circles).  The dashed line shows the contribution from the blackbody component.  For the optical/UV, we also show as squares the luminosity obtained when adopting a host extinction of ${E(B-V) = 0.042}$, instead of the value of $E(B-V) = 0.2$ assumed by \citet{ghisellini1}.  Both cases assume a Galactic extinction of $E(B-V) = 0.14$.  The stars show the values from \citet{campana}, which are uncorrected for absorption.}
\label{fig:SED}
\end{figure}

The total contribution to the XRT light curve can be separated into the contributions from blackbody and free-free processes, shown respectively by the dashed and dash-dotted lines in Fig.~\ref{fig:lightcurves}.  As expected, we find that the early emission is dominated by free-free emission, and the late emission is dominated by blackbody emission. The strength of the blackbody emission relative to the total emission is significantly lower in our model compared to \citet{campana}, although the time when the blackbody flux peaks is similar.  A similar conclusion can be drawn from Fig.~\ref{fig:SED}, where we compare our model spectrum at 2400\,s to the spectrum of \citet{ghisellini1}, who independently re-analysed the X-ray data of \citet{campana}.  This discrepancy could be due to absorption effects, since the $0.1\,$keV blackbody component is rather sensitive to the assumed absorption, whereas the harder free-free component is less affected by it.  The observed X-ray flux scales as $F_\nu \propto e^{-\tau_{\rm X}}$, where the X-ray absorption optical depth is  $\tau_{\rm X} \approx 2 (N_{\rm H}/10^{22}\,\text{cm}^2) (h\nu/1\,\text{keV})^{-2.4}$ (\citealt{kembhavi}, see also \citealt{morrison,balucinskachurch,wilms}), and $N_{\rm H}$ is the hydrogen absorption column.  \citet{campana} adopted a host galaxy column density of $N_{\rm H} = 6\times 10^{21}\,\text{cm}^{-2}$, but even a modest change to a value of $N_{\rm H} = 5\times 10^{21}\,\text{cm}^{-2}$ would reduce the flux by a factor of $\sim30$ at $0.3\,$keV, while the flux above 1\,keV would remain nearly unaffected.  Because of this extreme sensitivity to absorption (and because we do not wish to add additional freedom to our model), we do not concern ourselves with reproducing the exact features of the blackbody component reported by \citet{campana}, instead focusing on fitting the total XRT flux, which is more robust.

However, regardless of the assumed absorption, our model cannot explain the presence of the observed blackbody component from such early times.  In our model, unavoidably, blackbody emission is not present until $t>t_{\rm eq} \approx 2000\,$s, because at times $t<t_{\rm eq}$, the deeper layers of the ejecta where the gas and radiation are in equilibrium have not been revealed yet. A possible solution to this issue could be to consider non-homogenous or non-spherically symmetric ejecta.  In that case, different parts of the ejecta could have different velocities and temperatures, and some slower parts of the ejecta might be in thermal equilibrium from the beginning, so that blackbody emission could be present from early times.   We will reserve the study of the non-spherical case for future work.

Despite the challenges described above, we point out that the blackbody component's temperature evolution in our model, as $T_{\rm BB} \propto t^{-0.3}$, is roughly consistent with the slow decline in temperature reported in \citet{kaneko}.  Although they did not provide the luminosity of the blackbody component as a function of time for their fits, they claim that on average, the blackbody component contributed $0.13$ per cent of the total flux in the 0.5--150\,keV range during the prompt emission phase, which is similar to the value of $0.10$ per cent obtained in our model by averaging the 0.5--150\,keV flux over the first 3000\,s.  The analysis of the blackbody component by \citet{kaneko} therefore seems somewhat more consistent with our results.

\begin{figure}
\begin{center}
\includegraphics[width=\columnwidth]{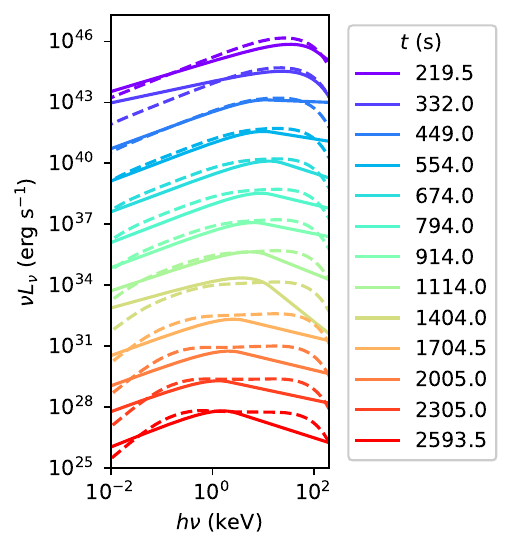}
\end{center}
\caption{Time evolution of the spectrum of the free-free component in our model (dashed lines), compared to the Band function spectral fits reported by \citet{toma} (solid lines).  The colours indicate different observer times, as shown in the legend.  For visual clarity, each subsequent spectrum after the first has been divided by an additional factor of 30.}
\label{fig:comparedspectra}
\end{figure}

Compared to the XRT light curve, the agreement with the BAT light curve is not as good.  Although our model reproduces the peak luminosity and time-scale of the BAT emission reasonably well, the shape of the light curve is significantly different from the data of \citet{toma}.  In particular, our model light curve is flatter near the peak, and has a steeper drop.  The difference in shape arises due to the difference in the spectrum assumed by \citet{toma} compared to the spectrum in our model.  \citet{toma} used a Band-function model for the spectrum, which is basically a broken-power law described by a single characteristic energy, and they found that the power-law above the break was typically steep.  On the other hand, as discussed above and shown in Fig.~\ref{fig:llgrbspectrumexample} (see papers I and II for further details), our model spectrum is essentially a free-free spectrum smeared over a range of temperatures, and it has two characteristic energy scales--we call them $kT_{\rm obs,min}$ and $kT_{\rm obs,max}$--which set the lower and upper bounds of this range.  In our model, above $h\nu > kT_{\rm obs,min}$, the spectrum is flatter than in \citet{toma} when $h\nu < kT_{\rm obs,max}$, and steeper when $h\nu > kT_{\rm obs,max}$.  This is illustrated in Fig.~\ref{fig:comparedspectra}, which shows a comparison between our model spectrum and theirs at several different times.  We see that at low energies (up to the first break in our model), the spectra are in good agreement. But at higher energies, the steep power-law in \citet{toma} is effectively replaced a flat power-law and exponential cutoff in our model. Therefore, in the model of \citet{toma}, the BAT light curve immediately starts to decline as a steep power-law once the break energy drops below the BAT band.  In contrast, in our model, due to the much flatter power-law between the first break at $kT_{\rm obs,min}$ and the second break at $kT_{\rm obs,max}$, when $kT_{\rm obs,min}$ drops below the BAT band, the BAT light curve initially has a much flatter decay, and it only starts to decline steeply once $kT_{\rm obs,max}$ drops below the BAT band.

We note that since GRB 060218 is rather soft, the photon statistics in BAT may not be sufficiently robust to distinguish between these two spectral models in practice (T. Sakamoto, private communication). We encourage future studies of $ll$GRBs to consider a broken power-law $+$ exponential cutoff model, which is physically motivated by shock breakout theory, in addition to employing the usual Band-function or cutoff power-law fits.

We also show the early UV/optical light curves in the inset of Fig.~\ref{fig:lightcurves}. Here, we have dereddened the data of \citet{campana} using $E(B-V) = 0.042$ for the host galaxy, as discussed in Section~\ref{sec:observations}. For this extinction, our model roughly reproduces the optical flux over the first $\sim 2000\,$s of the burst.  Moreover, Fig.~\ref{fig:SED} shows that the model spectrum roughly agrees with observations in the optical even at 2400\,s (comparing the squares to the solid line), although the spectral slope differs somewhat.  Considering that we only attempted to fit the prompt X-ray emission, the agreement with the optical data to within a factor of a few, despite the fact that the optical and XRT bands are separated by a factor of 100 in frequency, is remarkable. However, an issue with our model is that the optical emission decreases after a few thousand seconds, once the self-absorption frequency increases past the optical band.  In contrast, the observed optical emission continues to rise up to a peak at $0.5$\,d.  Another source of optical emission, such as the cooling envelope scenario discussed in Section~\ref{sec:opticalpeak}, would be needed to explain this peak.  The model used here only applies to the early planar phase of evolution, and does not include this.

\subsection{Constraints on the progenitor and explosion properties}
\label{sec:interpretation}

As discussed in Section~\ref{sec:results}, the six spectral parameters overconstrain the four shock breakout parameters $(M_{\rm env},R_{\rm env},E_0,n)$, and must therefore be checked for self-consistency.  Our approach is to explore a range of potential breakout parameters, and search for a region of this parameter space where the spectral parameters are reproduced to within some tolerance (we will consider $\alpha$ within $\pm0.5$ and other parameters within a factor of a few to be acceptable).  Based on the results of Paper I, we focus here on small values of $n \approx 0$, which are more favourable for INERT breakout, and greatly simplify the calculations.  A more detailed analysis which allows $n$ to vary is presented in Appendix~\ref{sec:appendixa}, where we also discuss the robustness of our findings.

To get an idea of the values of $E_0$, $M_{\rm env}$, and $R_{\rm env}$, it is simplest to start by considering only three of the spectral parameters: $L_{\rm bo}$, $t_{\rm bo}$, and $t_{\rm lc}$.  These three parameters have a straightforward power-law dependence on the breakout parameters, unlike $T_{\rm obs,bo}$, $t_{\rm eq}$, and $\alpha$, which depend non-trivially on $E_0$, $M_{\rm env}$, and $n$.  Adopting $n=0$, we can use the results in Section 3 of Paper I to obtain:
\begin{equation}
\label{E0}
    E_0 \approx 9 \times 10^{50}\,\text{erg} \left(\dfrac{t_{\rm bo}}{100\,\text{s}}\right)^{-1} \left(\dfrac{t_{\rm lc}}{1000\,\text{s}}\right)^{3}
\end{equation}
\begin{equation}
\label{Menv}
    M_{\rm env} \approx 0.8\,\mathrm{M}_\odot \left(\dfrac{L_{\rm bo}}{10^{46}\,\text{erg}\,\text{s}^{-1}}\right)^{-2}\left(\dfrac{t_{\rm bo}}{100\,\text{s}}\right)^{-3} \left(\dfrac{t_{\rm lc}}{1000\,\text{s}}\right)^{7} 
\end{equation}
\begin{equation}
\label{Renv}
    R_{\rm env} \approx 3\times 10^{13}\,\text{cm} \left(\dfrac{t_{\rm lc}}{1000\,\text{s}}\right) .
\end{equation}
For the desired values $t_{\rm lc} = 3000\,$s, $t_{\rm bo} = 100\,$s, and ${L_{\rm bo} = 4 \times 10^{47}\,\text{erg}\,\text{s}^{-1}}$, we obtain $E_0 \approx 2.4 \times 10^{51}\,$erg, $M_{\rm env} \approx 1 \mathrm{M}_\odot$, and $R_{\rm env} \approx 9 \times 10^{13}\,$cm.  Apart from the radius, these values seem unreasonable, as $M_{\rm env}$ is much larger than the low mass inferred from the optical peak, and $E_0$ is comparable to the entire supernova energy.  However, we note that these values are extraordinarily sensitive to $t_{\rm lc}$.  If the value of $t_{\rm lc}$ were 50 per cent smaller, we would instead have $E_{0} \approx 3 \times 10^{50}\,$erg, $M_{\rm env} \approx 8\times 10^{-3} \mathrm{M}_\odot$, and $R_{\rm env} \approx 5 \times 10^{13}\,$cm, which is comparable to the values obtained in past studies \citep[e.g.,][]{nakar,ic16}.

As noted above, we are interested here only in a rough solution which reproduces the spectral parameters to within a factor of a few.  If $L_{\rm bo}$, $t_{\rm bo}$, and $t_{\rm lc}$ are allowed to vary between $\frac{1}{3}$ and $3$ times their desired values in equations~\ref{E0}--\ref{Renv}, we obtain a reasonable constraint on the radius, $0.3 \la R_{\rm env} \la 3$, only a lower limit on the energy, ${E_0 \ga 3 \times 10^{49}\,\text{erg}}$, and no meaningful constraint on $M_{\rm env}$.  However, since we are interested in solutions compatible with extended low-mass envelopes, we will suppose for now that values of $M_{\rm env} \la 1 M_{\odot}$ are sensible.

Now, let us see if we can leverage the other spectral parameters to improve these constraints.  First, consider $T_{\rm obs,bo}$, whose value is extremely sensitive to the shock velocity at breakout, $v_{\rm bo}$, as shown in Paper I \citep[see also, e.g.,][]{ns}.  For values of $T_{\rm obs,bo}$ within the range $3\,\text{keV} \la kT_{\rm obs,bo} \la 30\,$keV (i.e., within a factor of 3 of the desired value of 9\,keV), the implied breakout velocity is in the range $0.1 \la v_{\rm bo}/c \la 0.2$ (see Section 4 of Paper I).  By using the fact that $v_{\rm bo} \propto (E_0/M_{\rm env})^{1/2}$ for $n=0$, we can convert the limit on $v_{\rm bo}$ to a restriction on the value of $E_0$, namely $3 M_{\rm env,-1} \la E_{\rm 0,50} \la 10 M_{\rm env,-1}$ where $E_{\rm 0,50} = E_0/(10^{50}\,\text{erg})$ and $M_{\rm env,-1} = M_{\rm env}/(0.1\,\mathrm{M}_\odot)$.

Next, we investigate what we can learn from $t_{\rm eq}$.  In particular, we will consider the ratio $t_{\rm eq}/t_{\rm bo}$, which has a value of $\approx 20$ for our nominal spectral parameters. This ratio, like $T_{\rm obs,bo}$, also depends mainly on $v_{\rm bo}$.  To be more specific, it depends only on the value of a dimensionless thermalization parameter, $\eta_{\rm bo}$,\footnote{For the definition of $\eta_{\rm bo}$ and further discussion of its importance, see Paper I and \citet{ns}.} which is strong function of the shock velocity and a very weak function of the breakout density, $\eta_{\rm bo}\propto \rho_{\rm bo}^{-1/8} v_{\rm bo}^{15/4}.$  Now, if both $t_{\rm eq}$ and $t_{\rm bo}$ are allowed to vary by a factor of 3, the ratio $t_{\rm eq}/t_{\rm bo}$ could vary by a factor of about 10.  Considering values in the range $2 \la t_{\rm eq}/t_{\rm bo} \la 200$, and following the procedure in Section 2 of Paper I, we find that $\eta_{\rm bo}$ lies in the range $2 \la \eta_{\rm bo} \la 10$.  This is then converted to a constraint on $E_0$ by using $\eta_{\rm{bo}} \approx 2.3 E_{0,50}^{15/8} M_{\rm{env,}-1}^{-2} R_{\rm{env,}14}^{3/8}$ for $n=0$ (Paper I), where $R_{\rm env,14}=R_{\rm env}/10^{14}\,$cm.  Neglecting the weak dependence of $\eta_{\rm bo}$ on the envelope density $\rho_{\rm env} \propto M_{\rm env} R_{\rm env}^{-3}$, we find $1 M_{\rm env,-1} \la E_{\rm 0,50} \la 3 M_{\rm env,-1}$.  Comparing this to the constraint obtained from the temperature, we see that they are marginally consistent, and together imply $E_{\rm 0,50} \sim 3 M_{\rm env,-1}$.  With this new constraint, the permissible range of values for $E_0$ and $M_{\rm env}$ is sharpened to $3 \la E_{\rm 0,50} \la 30$ and $1 \la M_{\rm env,-1} \la 10$.

Going a step further, this suggests that $\eta_{\rm bo}$ should be near the upper end of the allowed range, and likewise for the ratio $t_{\rm eq}/t_{\rm bo}$.  To achieve this, $t_{\rm eq}$ should be somewhat larger than the nominal value, and $t_{\rm bo}$ somewhat smaller.  If we suppose $t_{\rm bo} \sim 30\,$s, instead of $100$\,s as suggested from our spectral model, then the relation ${t_{\rm bo} \approx 3400\,\text{s}\,E_{0,50}^{-1} R_{\rm{env,}14}^3}$ (Paper I) implies ${E_{\rm 0,50} \sim 100 R_{\rm env,14}^3}$.  When combined with the relation ${E_{\rm 0,50}\sim 3 M_{\rm env,-1}}$ inferred above, this becomes a constraint on the breakout density, $\rho_{\rm bo} \sim 6\times 10^{-10}\text{g}\,\text{cm}^{-3}$, where we also used $\rho_{\rm{bo}} \approx 1.6 \times 10^{-11}\,\text{g}\,\text{cm}^{-3} M_{\rm{env,}-1} R_{\rm{env,}14}^{-3}$ for $n=0$ (Paper I).  The additional constraint relating $E_0$ to $R_{\rm env}$ can be used to further refine the allowed range of radii to $0.3 \la R_{\rm env,14} \la 0.6$.

Lastly, we consider the remaining spectral parameter, $\alpha$, which by definition satisfies $\alpha = \ln(T_{\rm obs,bo}/T_{\rm eq})/\ln(t_{\rm eq}/t_{\rm bo})$.  Up to a logarithmic correction which we neglect here for simplicity, the value of $T_{\rm eq}$ can be estimated via $T_{\rm BB,bo} (t_{\rm eq}/t_{\rm bo})^{-1/3}$ (Paper I), where ${T_{\rm BB,bo} \approx (\rho_{\rm bo} v_{\rm bo}^2/a)^{1/4}}$ is the temperature that the breakout material would have if it were in thermal equilibrium.  For $t_{\rm eq}/t_{\rm bo} \sim 200$, $v_{\rm bo} \sim 0.1\,c$, and $\rho_{\rm bo} \sim 6\times 10^{-10}\text{g}\,\text{cm}^{-3}$ as above, we find $kT_{\rm eq} \sim 10\,$eV, leading to $\alpha \approx 1.1$.  This is in some tension with the desired value of $\alpha = 1.6$. However, we note that like $T_{\rm obs,bo}$ and $t_{\rm eq}$, the value of $\alpha$ is extremely sensitive to the precise value of the shock velocity.

To summarize, assuming $n=0$, we find that in order to roughly explain the spectral parameters derived in Section~\ref{sec:prompt}, the breakout parameters must lie within the range
\begin{equation}
\label{Rinequality}
    3 \times 10^{13}\,\text{cm} \la R_{\rm env} \la 6 \times 10^{13}\,\text{cm},
\end{equation}
\begin{equation}
\label{Minequality}
     10^{-1}\,\mathrm{M}_\odot \la M_{\rm env} \la 1 \,\mathrm{M}_\odot,
\end{equation}
and
\begin{equation}
\label{Einequality}
    3 \times 10^{50}\,\text{erg} \la E_0 \la 3 \times 10^{51}\,\text{erg},
\end{equation}
and in addition must satisfy the constraints
\begin{equation}
    \label{MvsE}
    \dfrac{E_0}{10^{50}\,\text{erg}} \sim 3 \left(\dfrac{M_{\rm env}}{0.1\,\mathrm{M}_\odot}\right)
\end{equation}
and 
\begin{equation}
    \label{MvsR}
    \dfrac{M_{\rm env}}{0.1\mathrm{M}_\odot} \sim 30 \left(\dfrac{R_{\rm env}}{10^{14}\,\text{cm}}\right)^3.
\end{equation}
In any model meeting these criteria, the density and shock velocity at the breakout location turn out to be roughly the same, with $v_{\rm bo} \sim 0.1\,c$ and $\rho_{\rm bo} \sim 6\times 10^{-10}\,\text{g}\,\text{cm}^{-3}$.  This ensures that the spectral parameters $t_{\rm bo}$, $t_{\rm eq}$, $T_{\rm obs,bo}$, and $\alpha$, are preserved for any choice of model, as all of these depend only on combinations of $\rho_{\rm bo} \propto M_{\rm env} R_{\rm env}^{-3}$ and $v_{\rm bo} \propto (E_0/M_{\rm env})^{1/2}$.  The remaining spectral parameters, $L_{\rm bo}$ and $t_{\rm lc}$, also depend on $R_{\rm env}$, but since $R_{\rm env}$ is confined to a narrow range (equation~\ref{Rinequality}), they also remain roughly constant in any allowed model.  As a consequence, the expected X-ray emission should be broadly similar provided the above conditions are met.  However, we prefer models towards the lower end of the ranges in equations~\ref{Rinequality}--\ref{Einequality}, because in that case the implied envelope properties are consistent with the values obtained from modelling the early optical peak of SN 2006aj \citep{nakar,ic16}.  The analysis of \citet{emery} also found a comparable blackbody radius of $2.3\times 10^{13}\,$cm.

In Appendix~\ref{sec:appendixa}, we perform a more careful numerical analysis to determine the preferred values for the breakout parameters.  The results are in good agreement with the rough estimates presented here.  Additionally, if $n$ is allowed to vary, we find that the spectral parameters can be reproduced with greater accuracy for $n \approx 0.5\text{--}1$.  The mass, radius, and energy inferred in the higher $n$ case are not significantly different from the values estimated here for the $n=0$ case.  These low values of $n$ justify our assumption of $\lambda=1.5$ in Section~\ref{sec:results}.

Although our results in this section only roughly agree with Section~\ref{sec:results}, we emphasize that since the four shock breakout parameters are overconstrained by the six spectral parameters, there is not guaranteed to be any solution reproducing the spectral parameters, let alone a degenerate solution.  This is especially true in light of the very sensitive dependence of the spectral parameters on $v_{\rm bo}$.  Yet, we none the less found a degenerate solution, even after removing freedom by fixing one of breakout parameters, $n$.  The fact that there are many choices of breakout parameters which produce spectral parameters roughly comparable to the desired values underscores the overall plausibility of the shock breakout interpretation. 

To conclude our discussion, let us consider how the results of this section compare with previous works.  First of all, despite the fact that we only modelled the prompt X-rays, our overall results are consistent with the presence of a low-mass extended envelope surrounding the star, with properties similar to those inferred from modelling the optical peak at half a day.  The energy deposited in the envelope, the shock velocity, and the radius of the envelope we obtain are consistent with the values reported by \citet{nakar}. However, compared to his values and the rough estimates in Section~\ref{sec:theory}, the mass and density of the envelope inferred here are several times larger.  This discrepancy stems from the way that Paper I calculates the shock velocity.  Their blast wave model gives $v_{\rm bo} \approx 2.4 (E_0/M_{\rm env})^{1/2}$, which is somewhat larger than the usual rough estimate of $\approx (2E_0/M_{\rm env})^{1/2}$.  So, in our model, for the same energy, roughly three times more mass is needed to obtain the same shock velocity.  

Our model for the X-rays can also be compared with the picture of \citet{campana}, who found that the blackbody component is produced at a radius of $\approx 10^{12}\,$cm. 
In contrast, in our work, both the blackbody and non-blackbody emission components originate from the same radius ${R_{\rm env} \sim 3 \times 10^{13}\,\text{cm}}$. There are several effects that combine to explain this discrepancy.  First, as seen in Fig.~\ref{fig:SED}, the temperature of our blackbody component around the time of the X-ray peak is lower by a factor of $\sim 3$ compared to the value of 0.17\,keV given in \citet{campana}.  Second, our blackbody component is fainter than theirs by a factor of $\sim 10$ (see Fig.~\ref{fig:lightcurves}).  The final difference is more subtle. \citet{campana} computed the blackbody radius using the usual Stefan-Boltzmann law, which assumes that the spectrum is produced at the photosphere where $\tau=1$.  However, in our model, this is not the case because the opacity of the medium is dominated by electron scattering. In this case, the spectrum is formed at a larger optical depth $\tau > 1$, and the luminosity scales as $L \propto R^2 \tau^{-1} T^4$ \citep[see][for further discussion]{ns}.  In our model, the emission initially originates from an optical depth of $\tau \approx c/v_{\rm bo} \approx 10$, but the optical depth rises even during the planar phase since we include the logarithmic correction discussed by \citet[][see  Paper I for further discussion]{fs}. At times $t \sim t_{\rm eq}$ when the ejecta in thermal equilibrium have become visible and the observed spectrum is a blackbody, the emission originates from an optical depth of $\tau \sim 100$.  Taken together, the reduction in temperature and luminosity and the increase in optical depth due to electron scattering imply that the radius $R \propto L^{1/2} \tau^{1/2} T^{-2}$ is increased by a factor of $(0.1)^{1/2} 100^{1/2} (1/3)^{-2} \sim 30$ in our model compared to \citet{campana}, giving a value of $R \sim 3\times 10^{13}\,$cm, consistent with the results of this section.

\section{Discussion and conclusions}
\label{sec:conclusions}

After reviewing the observations and previous theoretical treatments of the well-known $ll$GRB event GRB 060218 (Section~\ref{sec:overview}), we robustly concluded that if this event was powered by shock breakout, the shock must have been relatively fast ($\approx 0.1\,c$), the breakout radius must have been relatively large ($10^{13}$--$10^{14}$\,cm), and the breakout must have been out of thermal equilibrium (Section~\ref{sec:theory}).  We identified three key features which are hard to explain with present shock breakout models: the fast evolution of the peak energy, the simultaneous existence of blackbody and power-law components in the spectrum, and the appearance of bright optical emission on time-scales shorter than $\la 1000\,$s.  We then proposed a revised shock breakout model to address these shortcomings, based on our own recent work in a separate paper (Paper I).  

The model considers shock breakout from an extended low-mass envelope, similar to the setup in \citet{nakar}, but includes the effects of rapid thermalization and light arrival time introduced in \citet{fs} and Paper I, and supposes that the breakout is in a regime where the gas and radiation are initially out of equilibrium, but  thermalized ejecta are revealed on a time-scale faster than $R_{\rm env}/c$.  In this case, as explained in Section~\ref{sec:physicalpicture}, the rapid evolution of the spectrum from a broad free-free spectrum to a blackbody spectrum in less than the light-crossing time causes these components to be blended together in the spectrum, as shown in Fig.~\ref{fig:llgrbspectrumexample}.  At peak bolometric light, the X-ray spectrum consists of a blackbody hump at $\sim 0.1\,$keV and a power-law tail extending to higher energies, similar to the spectrum seen in GRB 060218.  At later times, the blackbody component dominates, in agreement with \citet{campana}.  The conditions required to achieve this type of spectrum are consistent with expectations for $ll$GRBs (Paper I). Interestingly, we find that throughout the region of parameter space where this behaviour occurs, the expected peak energy is in the \textit{Swift} XRT band (Fig.~\ref{fig:temperaturemap}), making high-cadence soft X-ray missions like \textit{Swift} or \textit{Einstein Probe} uniquely well-suited to studying this type of unusual shock breakout.

In Section~\ref{sec:grb060218}, we applied the results of Paper I to build a simple spectral model for GRB 060218, based on the assumption that it was produced by a shock breakout in this newly realized regime.  As shown in Fig.~\ref{fig:lightcurves}, the model results in an excellent fit to the XRT light curve, and a satisfactory fit to the BAT light curve and the prompt \textit{V}-band light curve.  From this modelling, we infer an equilibrium time of $t_{\rm eq} \approx 2000\,$s and a light-crossing time of $t_{\rm lc}\approx 3000\,$s, and also constrain the properties of the breakout shell which dominates the early emission, finding that this shell emitted a luminosity of ${L_{\rm bo} \approx  4 \times 10^{47}\,\text{erg}\,\text{s}^{-1}}$ with an observed temperature of ${T_{\rm obs,bo} \approx 9\,\text{keV}}$, and it had a diffusion time of $t_{\rm bo} \approx 100\,$s.  We also showed that these values are self-consistent, in the sense that they indeed imply a breakout in the desired regime. 
 
Despite the model's successes, there are still several features of GRB 060218 which are difficult to interpret.  In particular, we are unable to explain why the $\approx 0.1\,$keV blackbody component is present from such early times, or how the ejecta is able to cool down to a temperature of $\approx 1\,$eV by 0.5\,d, as is required to explain the optical observations at late times.  These discrepancies are summed up by the statement that while our model does a good job explaining the behaviour of the hottest gas, at both early and late times there seems to be evidence for material that is colder than what our model predicts.  Furthermore, there is also evidence from the radio indicating a small amount of material travelling faster than what our model predicts.  We speculate that these problems could be alleviated by relaxing our simplifying assumptions of spherical symmetry and homogeneity of the ejecta. The strong polarization of SN 2006aj reported by \citet{gorosabel} further supports an aspherical interpretation.

Our modelling implies an envelope mass of $M_{\rm env} \sim 0.1\,\mathrm{M}_\odot$, an envelope radius of $R_{\rm env} \sim 3 \times 10^{13}\,$cm, and an energy of ${E_0 \sim 3 \times 10^{50}\,\text{erg}}$ deposited in the envelope, although there is considerable degeneracy as discussed in Section~\ref{sec:interpretation}.  Additionally, we constrain the density profile near the shock breakout location to be relatively flat, i.e. $\rho \propto (R_{\rm env}-r)^n$ with with $n=0$--1.  We stress that in our model, these values were obtained solely from considering the prompt X-ray emission of GRB 060218, yet they are none the less in reasonable agreement with the values obtained by other authors who modelled the early optical peak in SN 2006aj \citep{nakar,ic16}.  

The mass and radius of the extended material in our $ll$GRB model can be compared with the expectations for other stripped envelope SNe of Types Ibn \citep[see][and references therein]{hosseinzadeh17} and Icn \citep[see][and references therein]{pellegrino22}, which also show evidence of interaction with a dense CSM. While it is important to stress that there is considerable diversity in both of these classes, there are some emerging trends.  Recent models of Ibn SNe \citep[e.g.,][]{moriyamaeda,dessart,maedamoriya} suggest a low SN ejecta mass interacting with a CSM mass of $\sim 0.01$--$1\,\mathrm{M}_\odot$ at radii of $\sim 10^{15}\,$cm.  Meanwhile, for their sample of Icn SNe, \citet{pellegrino22} estimated a CSM mass of $0.1$--$0.5\,\mathrm{M}_\odot$ interacting with an ejecta mass of $\la 2\,\mathrm{M}_\odot$ travelling at $\sim 10^9\,\text{cm}\,\text{s}^{-1}$, which at $\sim 10\,$d implies a radius of $\sim 10^{15}\,$cm for the interacting shell.  Although the mass in both cases is comparable to what we obtain for $ll$GRBs, it seems that the CSM in Ibn and Icn SNe must extend to significantly larger radii, and furthermore \citet{maedamoriya} found the need for a steep density profile $\rho \propto r^{-3}$, whereas our model requires a flatter profile with most of the mass concentrated near $R_{\rm env}$. Taken together, the results for $ll$GRBs, Ibn SNe, and Icn SNe emphasize the need for a stellar evolution channel for Type Ibc SNe which can strip the SN progenitor down to a few solar masses while also depositing up to $1\,\mathrm{M}_\odot$ of material somewhere between $10^{13}$ and $10^{15}\,$cm.

It is interesting to note that our model requires a relatively large energy of several $10^{50}\,$ ergs to be deposited into the extended envelope, which is a substantial fraction of the energy of SN 2006aj, $2\times 10^{51}\,$erg \citep{mazzali}.  This may be difficult to accomplish if the envelope is shocked by the fast outer layers of the supernova, as the fastest supernova ejecta typically carry only a small fraction of the total energy \citep[e.g.,][]{tan}.  Possibly, this indicates that the energy was deposited by a choked jet, as was previously suggested by \citet{nakar}.  The jet could be choked in the extended envelope, or in the outer layers of the star, as long as it is able to transfer $\ga 10^{50}\,$erg into the envelope.  The case of a jet choked in an extended CSM was recently investigated numerically by \citet{suzuki}, who found that a typical GRB jet suffocated by a CSM with a mass of $0.1\,\mathrm{M}_\odot$ and a radius of $3\times 10^{12}$--$3\times 10^{13}\,$cm could roughly reproduce the early optical peak of SN 2006aj.  If a choked jet is responsible for exploding the envelope, it is unlikely that the outflow would manage to completely spherize before breakout \citep[see, e.g.,][]{inp}, so an axisymmetric shock breakout would naturally be expected.  This scenario is especially interesting because in an axisymmetric breakout, the breakout properties, including the initial temperature, vary as a function of latitude along the breakout surface \citep[as discussed by, e.g.,][]{matzner,il}, which provides a promising way to solve the problem of cooler-than-expected material in our model.  Furthermore, as the breakout duration can be prolonged by a factor of up to $c/v_{\rm bo}$ in the axisymmetric case \citep{il}, we expect that the conditions for an INERT breakout will be easier to achieve for asymmetric shocks.  We plan to further explore the axisymmetric shock breakout scenario in future work.

Although there seems to be a growing consensus that the presence of a low-mass envelope or CSM can help to explain $ll$GRBs, the origin of this extended material remains unknown.  In Type IIb SNe, the star is thought to be inflated by interaction with a binary companion \citep[e.g.,][]{podsiadlowski92,podsiadlowski93,eldridge08,stancliffe09,yoon10,claeys11,smith11}, but it is not clear if this scenario is applicable to H-poor progenitors.  Binary evolution models of stripped Type Ibc progenitors \citep[e.g.,][]{yoon10,tauris13,tauris15,laplace20} typically do not have enough mass present at large radii to explain $ll$GRBs \citep[see, however,][]{wufuller22b}.  Therefore, an episode of enhanced mass-loss in the days to months before explosion may be required to produce the desired properties. Several promising mass-loss mechanisms have been suggested, including wave-driven mass-loss \citep{quataert,mcleysoker,shiode,fuller17,fullerro18,ouchimaeda19,leungfuller20,leungfuller21a,leungfuller21b,matzner21,wufuller22a}, pulsational pair instability-driven mass-loss \citep{woosley17,leungnomoto19,marchant19,woosley19,leungblinnikov20,renzo20}, and interaction or merger with a compact object companion \citep{chevalier12,danielisoker,schroder20}.  However, in the case of GRB 060218, each of these has issues.  In the most up-to-date wave-drive mass-loss models \citep[e.g.,][see their Figure 12]{leungfuller21b}, the predicted CSM radius is comparable to our value of $\sim 3 \times 10^{13}\,$cm, but the ejected mass is typically $< 10^{-2}\,\mathrm{M}_\odot$ for H-poor stars, less than the $\sim 0.1\,\mathrm{M}_\odot$ our model requires.  In the pulsational pair instability case, the  possible CSM masses and radii span a wide range which could accommodate our desired values, but the predicted pre-explosion stellar mass is $\sim 40\,M_{\odot}$ \citep[e.g.,][]{renzo20}, much higher than the ejecta mass of $2 \mathrm{M}_\odot$ inferred for SN 2006aj \citep{mazzali}. The merger scenario may be problematic for H-stripped progenitors, since a compact object merging with a He star is expected to power a successful jet and produce a transient more akin to a typical long GRB \citep[e.g.,][]{fryerwoosley,zhangfryer}, although it is possible that the star could be tidally disrupted and accreted on to the compact object to launch a disc wind instead \citep[see][for further discussion]{metzger22}.    Alternatively, \citet{wufuller22b} recently proposed that stars with a He core mass between $2.5\,\mathrm{M}_\odot$ and $3\,\mathrm{M}_\odot$ can undergo extreme binary mass transfer, resulting in a CSM with a mass of $0.01$--$1\,\mathrm{M}_\odot$ and a radius of $10^{13}$--$10^{16}\,$cm. These values are fully consistent with the range in our equations~\ref{Rinequality}--\ref{Minequality}, and with the low supernova mass inferred by \citet{mazzali}.  However, in this picture it is not clear how to explain the lack of He lines in SN 2006aj.

Intriguingly, a growing population of SNe \citep{mauerhan,ofek,strotjohann,jg22,fransson22,eliasrosa24,dong24}, including some Type Ibn SNe \citep{pastorello,brennan}, have been linked with SN precursor events.  The physics of SN precursors and the potential connection between precursor eruptions and the dense CSM indicated for interacting SNe is a rapidly developing field \citep[e.g.,][]{kuriyama20,leungfuller20,kuriyama21,linialfuller,tsuna21,ko,matsumotometzger,tsuna23a,tsuna23b,tsuna24a,tsuna24b}.  The precursor emission, if detected, can provide crucial independent constraints on the CSM mass and radius.  Observing a precursor in a future nearby $ll$GRB would therefore be enormously valuable for identifying the progenitor channels and mass-loss processes relevant to these peculiar transients.

Regardless of the formation mechanism of the envelope, it is clear that in our scenario, a significant fraction of broad-lined Type Ic SNe must possess an extended envelope in order to explain the observed $ll$GRB rate. \citet{guetta} estimated that $ll$GRBs occur in about 1--9 per cent of Type Ibc SNe (assuming isotropic emission), while standard long GRBs occur in about 0.4--3 per cent of Type Ibc SNe (assuming a beaming factor of 75--500).  Although their $ll$GRB rate is for all events less luminous than $10^{49}\,\text{erg}\,\text{s}^{-1}$, they find that the rate of GRB 060218-like events specifically is comparable, in agreement with \citet{soderberg}.  Meanwhile, \citet{guetta} estimated that broad-lined Type Ibc SNe account for $\la 10$ per cent of Type Ibc SNe, which suggests that 10--100 per cent of broad-lined Type Ibc SNe produce $ll$GRBs, and 4--30 per cent of them produce typical long GRBs.  In the context of our model for GRB 060218, several interesting conclusions can be drawn from these results.  Let us suppose that a fraction $f_{\rm env}$ of broad-lined Type Ic SN progenitors have extended envelopes, and a fraction $f_{\rm jet}$ have jets (whether choked or successful) which are powerful enough to produce a GRB (including $ll$GRBs).  What can we say about $f_{\rm env}$ and $f_{\rm jet}$ from the available data? First, if extended envelopes are required to explain $ll$GRBs, then we must have $f_{\rm env} \ga 0.1$.  Order-unity values of $f_{\rm env}$ can probably already be ruled out by the fact that ZTF has seen broad-lined Type Ic events with no evidence for an early optical peak.  However, events with an early peak are also observed occasionally, and interestingly, as discussed in Section~\ref{sec:history}, one ZTF event with an early optical peak, SN 2020bvc, also had an afterglow resembling the afterglow of GRB 060218 \citep{ho}.  Second, if both $ll$GRBs and standard GRBs require jets, then we would need $f_{\rm jet} \ga 0.1$ as well.  Placing an upper limit on $f_{\rm jet}$ is difficult from current observations, because faint $ll$GRB-like prompt emission is hard to rule out.  Finally, if the presence of an envelope is not correlated with the presence of a jet, then there would be four possible scenarios for broad-lined Type Ic SNe: typical long GRBs which require a jet, but no envelope, with a probability of $f_{\rm jet}(1-f_{\rm env})$; $ll$GRBs which require both a jet and an envelope, with probability of $f_{\rm jet}f_{\rm env}$; double-peaked SNe Ic without a GRB, with a probability of $f_{\rm env}(1-f_{\rm jet})$; and typical SNe Ic without an early optical peak or a GRB, with a probability of $(1-f_{\rm env})(1-f_{\rm jet})$.

In this simple picture, the observed ratio of $ll$GRBs to standard GRBs should be equal to the ratio of double-peaked to single-peaked broad-lined Type Ic SNe, with both ratios given by $f_{\rm env}(1-f_{\rm env})^{-1}$.  However, although there is still considerable uncertainty, a preference for small $f_{\rm env}$ seems to be in tension with the apparent abundance of $ll$GRBs compared to high-luminosity bursts.  We offer several possible resolutions to this: 1) standard GRBs might require rarer, more extreme jets than $ll$GRBs; 2) the possibility that standard GRBs and $ll$GRBs have comparable rates cannot be ruled out, if the beaming factor is large \citep{guetta}; 3) the presence of jets and extended envelopes might in fact be correlated, for example if they both arise preferentially in binaries; 4) low-mass envelopes will peak in the optical on short time-scales, and may be missed in surveys.    High-cadence optical surveys such as ZTF should eventually be able to constrain the fraction of broad-lined Type Ic SNe which have an early optical peak on a time-scale of a day, which will at least put a strict lower limit on $f_{\rm env}$ and greatly help with clarifying this progenitor scenario.

Despite the many perplexing features of $ll$GRBs, the future of $ll$GRB science is bright.  Our understanding of the spectral evolution and bolometric luminosity of $ll$GRBs is currently held back considerably by the limited UV sensitivity of \textit{Swift} and the lack of spectral coverage around tens of eV.  The upcoming UV mission \textit{UVEX} will provide higher UV sensitivity than \textit{Swift}, far-UV capabilities, and a target-of-opportunity response time of hours \citep{kulkarniUVEX}, which should grant us significantly more insight into the properties of the UV/optical peak in $ll$GRBs.  In addition, the upcoming \textit{ULTRASAT} mission offers an aperture similar to \textit{Swift}'s UVOT but a much larger field-of-view \citep{shvartzvald}.  \textit{ULTRASAT} is poised to discover a wide variety of UV-bright shock breakout transients, including $ll$GRBs, and will serve as an excellent testbed for our shock breakout predictions. Finally, the \textit{Einstein Probe} mission, which launched in January 2024, offers a significantly improved sensitivity and field of view in soft X-rays compared to \textit{Swift} \citep{yuan}, making it ideal for detecting soft, low-luminosity bursts.  We expect that the number of known $ll$GRBs will increase greatly over the coming years, and are hopeful that with a larger sample size will come a clearer understanding of these peculiar transients which have puzzled theorists for over 20 years. \\ \\

\section*{Acknowledgements}
We are indebted to T. Sakamoto for discussions which clarified our understanding of the dataset for GRB 060218 and greatly improved the paper.  We also thank T. Faran, E. Nakar, R. Sari, and K. Murase for fruitful discussions and comments.  Additionally, CI thanks A. Ho and M. Modjaz for helpful discussions about double-peaked SNe, and J. Fuller and S.-C. Leung for discussions about possible mass-loss mechanisms in Type Ibc SNe.  This work is supported by the JST FOREST Program (JPMJFR2136) and the JSPS Grant-in-Aid for Scientific Research (20H05639, 20H00158, 23H01169, 23H04900). 

\section*{Data Availability} 
%The inclusion of a Data Availability Statement is a requirement for articles published in MNRAS. Data Availability Statements provide a standardised format for readers to understand the availability of data underlying the research results described in the article. The statement may refer to original data generated in the course of the study or to third-party data analysed in the article. The statement should describe and provide means of access, where possible, by linking to the data or providing the required accession numbers for the relevant databases or DOIs.

The data underlying this article will be shared on reasonable request
to the corresponding authors.

%%%%%%%%%%%%%%%%%%%%%%%%%%%%%%%%%%%%%%%%%%%%%%%%%%

%%%%%%%%%%%%%%%%% APPENDICES %%%%%%%%%%%%%%%%%%%%%

%\appendix



\appendix

\section{Method of determining the progenitor and explosion properties}
\label{sec:appendixa}

In Section~\ref{sec:interpretation}, we provided some rough estimates for the properties of the progenitor and the explosion based on the simplifying assumption $n=0$.  Here, we conduct a numerical study to confirm these approximate findings, and verify that varying $n$ does not significantly affect our conclusions.  

We explore a grid of the breakout parameters $E_0$, $M_{\rm env}$, $R_{\rm env}$, with $\log_{10} E_{\rm 0,50}$ ranging from -0.4 to 1.6 in steps of 0.02, $\log_{10} M_{\rm env,-1}$ ranging from -1 to 1 in steps of 0.02, $\log_{10} R_{\rm env,14}$ ranging from -1 to 0 in steps of 0.25, and $n$ ranging from 0 to 1.5 in steps of 0.5.  The broad range of $E_0$ and $M_{\rm env}$ compared to $R_{\rm env}$ is chosen to accommodate the degeneracy anticipated from equations~\ref{Rinequality}--\ref{Einequality}.  The lower and upper bounds of $n=0$ and $n=1.5$ represent, respectively, the case where the breakout occurs before shock acceleration becomes significant (as discussed in Section 3 of Paper I), and the case of standard shock breakout from the edge of medium supported by convection \citep[e.g.,][]{ns}.  

For each point in the parameter grid, we calculate the values of the spectral parameters according to the procedure outlined in Section 3 of Paper I.  We note that the observed temperature depends on the degree of Comptonization, which is captured by a dimensionless parameter $\xi_{\rm bo}$ \citep[see, e.g.,][and Paper I]{ns,fs,margalit}.  Here we adopt the Comptonization model described in Section 4 of Paper I, but we will also explore below how the results change if $\xi_{\rm bo}$ differs from their assumption.

Next, we compare the values obtained from this procedure to the values of ${t_{\rm bo} = 100\,\text{s}}$, $t_{\rm eq} = 2000\,$s, $t_{\rm lc} = 3000\,$s, ${L_{\rm bo} = 4 \times 10^{47}\,\text{erg}\,\text{s}^{-1}}$, $kT_{\rm obs,bo} = 9\,$keV, and $\alpha=1.6$ obtained from our spectral model in Section~\ref{sec:prompt}, and compute the residual sum of squares (RSS)
\begin{align}
    \label{RSS}
    \text{RSS} & = \left[\log_{10} \left(\dfrac{t_{\rm bo}}{100\,\text{s}} \right)\right]^2 + \left[\log_{10} \left(\dfrac{t_{\rm eq}}{2000\,\text{s}} \right)\right]^2\nonumber \\ 
    & \quad + \left[\log_{10} \left(\dfrac{t_{\rm lc}}{3000\,\text{s}} \right)\right]^2   + \left[\log_{10} \left(\dfrac{L_{\rm bo}}{4 \times 10^{47}\,\text{erg}\,\text{s}^{-1}} \right)\right]^2 \nonumber \\
    & \quad + \left[\log_{10} \left(\dfrac{kT_{\rm obs,bo}}{9\,\text{keV}} \right)\right]^2   + (\alpha-1.6)^2
\end{align}
for each cell in the grid.  For each parameter besides $\alpha$, we take the residual of its logarithm.  This weighting gives preference to models with the desired $\alpha$, to account for the fact that $\alpha$ appears in  the exponent in our model and therefore most strongly influences the resulting spectrum.  To put it another way, changes in the parameters other than $\alpha$ by an order-unity multiplicative factor are acceptable in our model, but this is not the case for $\alpha$.

\begin{figure}
    \centering
    \includegraphics[width=\linewidth]{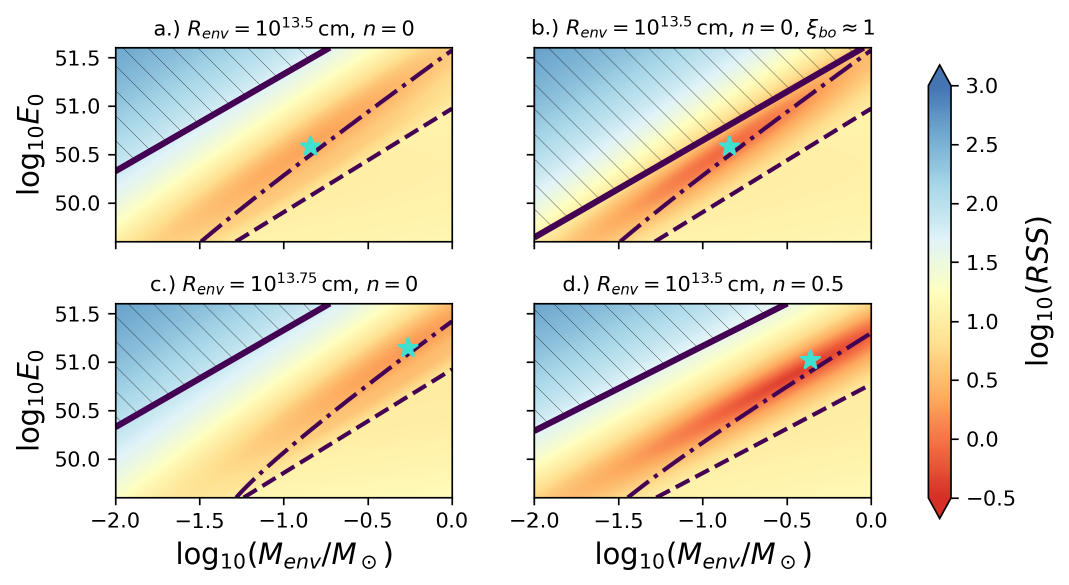}
    \caption{The RSS obtained by comparing the spectral parameters used to fit observations in Section~\ref{sec:results} to the values predicted by the shock breakout model of Paper I, throughout $M_{\rm env}$--$E_0$ parameter space for various choices of $R_{\rm env}$ and $n$.  The same axes are used in all cases.  The turquoise star indicates the model that minimizes the RSS.  As in Fig.~\ref{fig:temperaturemap}, the dashed line indicates $t_{\rm eq} = t_{\rm bo}$, while the dot-dashed line denotes $t_{\rm eq} = t_{\rm lc}$.  The region where pair creation becomes relevant ($kT \ga 50\,$keV) is hatched.  \textit{Panels:} a.) $R_{\rm env} = 10^{13.5}\,$cm and $n=0$, with the Comptonization model of Paper I. b.) $R_{\rm env} = 10^{13.5}\,$cm and $n=0$, but neglecting Comptonization up to the pair creation threshold. c.) As in panel a, but $R_{\rm env} = 10^{13.75}\,$cm.  d.) As in panel a, but $n=0.5$.}
    \label{fig:RSS}
\end{figure}

The results of this analysis are shown in Fig.~\ref{fig:RSS}, where we plot the RSS versus $M_{\rm env}$ and $E_0$ , fixing $R_{\rm env}$ and $n$, for several cases of interest.  Regions coloured in orange or red have $\text{RSS} \la 3$, indicating a satisfactory reproduction of the desired spectral parameters.  The cyan star indicates the location where the RSS is minimized.  In each case, we find that minimum RSS occurs very close the the boundary of the INERT breakout region, i.e. the dot-dashed line in the figure along which $t_{\rm eq} = t_{\rm lc}$.  In other words, we have $t_{\rm bo} \ll t_{\rm eq} \sim t_{\rm lc}$, consistent with our spectral model in Section~\ref{sec:prompt}.  The region of low RSS consistently overlaps the region between the dashed and dash-dotted lines, so an INERT breakout scenario seems plausible. 

The upper-left panel shows the case case considered in Section~\ref{sec:interpretation}, where $R_{\rm env} = 10^{13.5}\,$cm and $n=0$.  In this case, the best fit to the spectral parameters occurs for $M_{\rm env} = 0.14\,\mathrm{M}_\odot$ and ${E_0 = 4.0 \times 10^{50} \,\text{erg}}$, in good agreement with the values of $10^{-1}\,\mathrm{M}_\odot$ and $3 \times 10^{50}\,$erg suggested in Section~\ref{sec:interpretation}.  As expected, the region where the spectral parameters are reproduced is degenerate, and lies very nearly along the line $E_{\rm 0,50} = 3 M_{\rm env,-1}$.  A reasonable fit is obtained down to values of $E_0 \approx 10^{50}\,$erg and $M_{\rm env}\approx 3 \times 10^{-2}\,\mathrm{M}_\odot$, somewhat lower than the estimated limit in Section~\ref{sec:interpretation}.

In the upper-right panel of Fig.~\ref{fig:RSS}, we consider the effects of reduced Comptonization.  The flat low-energy spectral index of the non-blackbody X-ray emission \citep[e.g.,][]{toma} and the lack of an apparent Wien peak in the spectrum argue for a low degree of Comptonization in GRB 060218.  However, Paper I found that a modest amount of Comptonization is expected for temperatures relevant to $ll$GRBs, although they claim that it is not enough to significantly affect the spectrum.  To check whether their assumed Comptonization model (for which $\xi_{\rm bo} \sim 3$ when $kT_{\rm obs,bo} \sim 10\,$keV) affects the inferred envelope properties, we also consider a model where the Comptonization is artificially reduced by setting $\xi_{\rm bo} = 1$.  The best-fitting values for $M_{\rm env}$ and $E_0$ in this case turn out to be almost identical to before: $M_{\rm env} = 0.14\,\mathrm{M}_\odot$ and $E_0 = 3.8 \times 10^{50} \,$erg.  Interestingly, the fit is improved in this case, due to the somewhat higher values of $T_{\rm obs,bo}$ and $\alpha$.

We next investigate the effect of increasing the envelope radius to $R_{\rm env} = 10^{13.75}\,$cm in the lower-left panel.  As expected from Section~\ref{sec:interpretation}, there is again degeneracy along the line $E_{\rm 0,50} = 3 M_{\rm env,-1}$. The best-fitting mass and energy are shifted up to $M_{\rm env} = 0.55\,\mathrm{M}_\odot$ and $E_0 = 1.4 \times 10^{51} \,$erg, roughly consistent with the scaling $E_0 \propto M_{\rm env} \propto R_{\rm env}^3$ predicted in Section~\ref{sec:interpretation}. The quality of the fit is not significantly improved compared to the $R_{\rm env} = 10^{13.5}\,$cm case.  However, if the radius is further increased to $R_{\rm env} = 10^{14}\,$cm, we find that a good fit is no longer possible for the masses and energies we consider.  This reaffirms the upper limit of $R_{\rm env} \la 6 \times 10^{13}\,$cm suggested by equation~\ref{Rinequality}.  

Finally, in the lower-right panel we show the result of increasing $n$ from 0 to 0.5.  Increasing $n$ has a similar effect as increasing the radius since, for the same envelope mass, it reduces the density in the outer parts of the envelope.  To compensate the change in density and preserve the same spectral parameters, $M_{\rm env}$ must be increased, and $E_0$ must also be increased in proportion to $M_{\rm env}$ to preserve the same shock velocity.  As in the other cases, we find that there is a strip in the $M_{\rm env}$--$E_0$ plane where the spectral parameters can be reasonably reproduced.  However, compared to the $n=0$ case, this strip extends over an even larger range of $M_{\rm env}$ and $E_0$, with the fit being improved throughout the strip.  Additionally, the slope of the strip is slightly reduced: due to the effects of shock acceleration, $v_{\rm bo}$ scales slightly differently with $M_{\rm env}$ and $E_0$ for different $n$, so that lines of constant $v_{\rm bo}$ follow $E_0 \propto M_{\rm env}^{0.86}$ for $n=0.5$, instead of $E_0 \propto M_{\rm env}$ as in the other panels.  The best-fitting values in this case are $M_{\rm env} = 0.44\,\mathrm{M}_\odot$ and $E_0 = 1.1 \times 10^{51}\,$erg.  If $n$ is further increased to $n=1$, the fit remains equally good but the best-fitting values are shifted to lie outside of the range we consider.  For $n=1.5$, the fit becomes worse, and furthermore, there is barely any region where INERT breakout can occur within our parameter range. In addition, for $n=1.5$, the region where INERT breakout remains possible no longer overlaps the region where the spectral parameters can be reproduced.  For this reason, $n=1.5$ (or larger) is disfavoured for our model, but values of $n=0$, 0.5, or 1 seem plausible.

%%%%%%%%%%%%%%%%%%%%%%%%%%%%%%%%%%%%%%%%%%%%%%%%%%

% Don't change these lines
\bsp	% typesetting comment
\label{lastpage}
\end{document}